\documentclass{pasa}

\usepackage{graphicx}
\usepackage{epstopdf}
\usepackage{gensymb}
\usepackage{etoolbox}
\usepackage{float}

\makeatletter
\patchcmd{\ttlh@hang}{\parindent\z@}{\parindent\z@\leavevmode}{}{}
\patchcmd{\ttlh@hang}{\noindent}{}{}{}
\makeatother

\title[Pulsar Timing Arrays]{Pulsars probe the low-frequency gravitational sky: Pulsar Timing Arrays basics and recent results}

\author[Caterina Tiburzi]{Caterina Tiburzi$^{1,2}$\\
\affil{$^1$Max-Planck-Institut f\"{u}r Radioastronomie, Auf dem H\"{u}gel 69, 53121 Bonn, Germany}
\affil{$^2$Fakult\"{a}t f\"{u}r Physik, Universit\"{a}t Bielefeld, Postfach 100131, 33615 Bielefeld, Germany}
}

\jid{PASA}
\doi{10.1017/pas.\the\year.xxx}
\jyear{\the\year}

\usepackage{aas_macros}
\usepackage{hyperref}
\makeatletter
\newcommand\footnoteref[1]{\protected@xdef\@thefnmark{\ref{#1}}\@footnotemark}
\makeatother
\hypersetup{colorlinks,citecolor=blue,linkcolor=blue,urlcolor=blue}

\hypersetup{draft}

\DeclareRobustCommand{\VAN}[3]{#2} 

\begin{document}

\begin{frontmatter}
\maketitle

\begin{abstract}
 Pulsar Timing Array (PTA) experiments exploit the clock-like behaviour of an array of millisecond pulsars, with the goal of detecting low-frequency gravitational waves. PTA experiments have been in operation over the last decade, led by groups in Europe, Australia, and North America. These experiments use the most sensitive radio telescopes in the world, extremely precise pulsar timing models and sophisticated detection algorithms to increase the sensitivity of PTAs. No detection of gravitational waves has been made to date with this technique, but PTA upper limits already contributed to rule out some models of galaxy formation. Moreover, a new generation of radio telescopes, such as the Five hundred metre Aperture Spherical Telescope and, in particular, the Square Kilometre Array, will offer a significant improvement to the PTA sensitivity. In this article, we review the basic concepts of PTA experiments, and discuss the latest results from the established PTA collaborations. 
\end{abstract}

\end{frontmatter}

\section{Introduction}\label{sec:intro}

Pulsars are highly-magnetized, rapidly rotating neutron stars that convert their rotational kinetic energy into magnetic dipole radiation. Although their emission can extend to the entire electromagnetic spectrum, they are typically observed at radio wavelengths. \\
In a simplified picture, pulsar radio emission is generated in proximity of the magnetic poles, and forms radiation beams. If the magnetic and the spin axes are misaligned, then the two beams rotate with the neutron star and sweep through space. An observer whose line of sight crosses one or both of the beams, will observe a pulsed emission, the period of which corresponds to the pulsar's spin period (a general introduction to pulsar astronomy can be found, e.g., in \citealt{lk05}). Pulsars are often referred to as ``cosmic clocks'', because it is possible to predict the arrival time of each pulse at a telescope, sometimes with sub-microsecond precision \citep{dcl16,abb15,rhc16}, through the \textit{pulsar timing} technique (see Section~\ref{sec:timing}).\\

Due to their high densities, rapid rotations, strong magnetic fields, and high surface gravity, neutron stars are ideal laboratories for tests of nuclear physics \citep{lp04}, general relativity \citep{ksm+06} and alternative theories of gravity \citep{sck13} in extreme conditions not feasible in Earth-based laboratories (see also \citealt{sta03,ch08}). In the context of general relativity (GR) tests, Pulsar Timing Array (PTA, \citealt{fb90}) experiments are among the most exciting projects of the last decade. The primary aim of PTAs is the direct detection of low-frequency gravitational waves (GWs) \citep{rr95a,wl03a,shmv04}. \\

In this article, we describe the basic concepts and approaches of PTA experiments, and we review the recent results from the established PTA experiments. In Section~\ref{sec:gw}, we review the efforts to detect GWs in different parts of the spectrum. In Section~\ref{sec:timing}, we give an overview of the technique of ``pulsar timing'', used to interpret pulsar data for the purpose of PTAs. In Section~\ref{sec:sources}, we outline the potential sources of GWs at low frequencies, and in Section~\ref{sec:pta} we describe the basic concepts of PTA experiments. In Section~\ref{sec:results}, we summarize the latest results from the existing PTA experiments, and in Section~\ref{sec:future} we discuss the future prospects of PTAs, also considering the new radio astronomical facilities.

\section{The quest for gravitational waves}\label{sec:gw}

\subsection{Gravitational waves}
GWs were an early prediction of GR \citep{ein16}, and are a consequence of a small (i.e. linearisable) perturbation $h_{\mu,\nu}$ to an otherwise flat (or Minkowskian) metric $\eta_{\mu\nu}$ of space-time, produced by asymmetric and accelerated mass distributions:

\begin{equation}
 g_{\mu\nu} = \eta_{\mu\nu} + h_{\mu\nu}
\end{equation}

\noindent It is possible to demonstrate that the perturbation $h_{\mu\nu}$ propagates in the metric as a transverse wave at the speed of light, and in its propagation, it induces quadrupolar perturbations of space-time. Given a mass distribution (we recall that, in GR, the presence of a mass distribution curves the space-time), it is also possible to demonstrate that such perturbations $h_{\mu\nu}$ are generated if the second time derivative of the quadrupole mass moment $Q$ is not zero \citep{mag07}. \\

 The amplitude of GWs is typically expressed in terms of the dimensionless \textit{strain} $h$, i.e. the fractional change $\delta L$ induced by GWs over a distance $L$:

\begin{equation}
 h = \frac{\delta L}{L}.
\end{equation}

 \noindent GWs can have two polarizations, commonly referred to as ``plus'' and ``cross''. If a ``plus''-polarized GW propagates along the $z$ axis, then it will alternatively stretch and compress space-time along the $y$ and $x$ axes in the orthogonal direction. A ``cross''-polarized GW will have the same effect, although rotated by $45\degree$. \\

\begin{figure}
\begin{center}
 \includegraphics[scale=0.43]{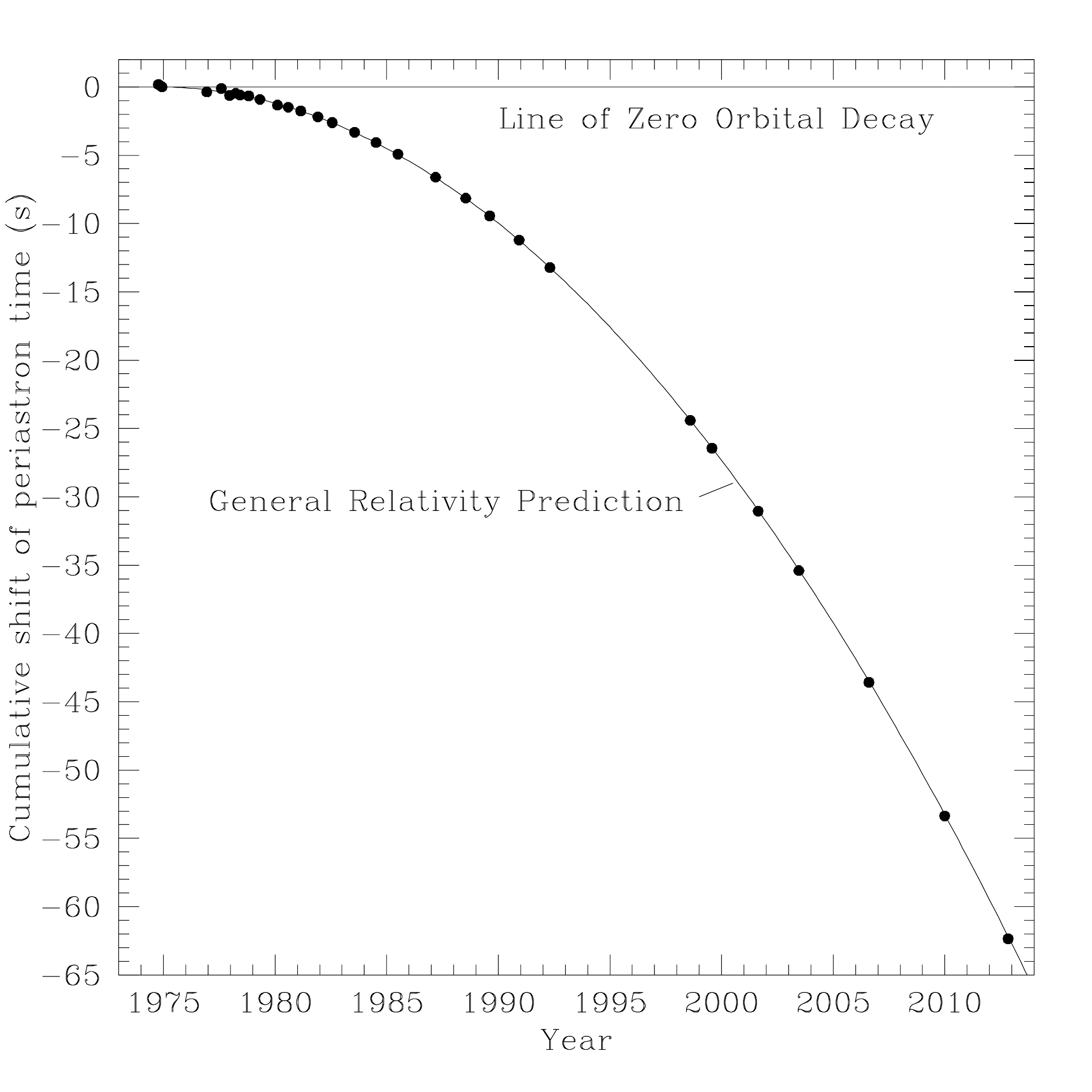}
\caption{Figure taken from \citet{wh16}\protect\footnotemark; Orbital phase shift of the double neutron star system that includes PSR\,B1913+16 versus time. The plot shows the perfect agreement between the observed orbital decay of the (black dots) and the prediction by GR (solid line).}
\label{fig:orbitaldecay1913}
\end{center}
\end{figure}

 Pulsar astronomy brought the first indirect confirmation of the existence of GWs, through observations of PSR\,B1913+16 \citep{ht74}. This object is a pulsar in a $\sim7.7$-hour orbit with another neutron star. By assuming the existence of GWs, GR can predict the rate of orbital decay that can be attributed to GW emission due to the orbital motion of the two neutron stars. The orbital decay of this binary system was found in agreement with the predictions, today to a precision greater than $99.5\%$ (\citealt{wt81, wnt10, wh16}, and see Figure~\ref{fig:orbitaldecay1913}).\\

\footnotetext{Figure 3 of \textit{Relativistic Measurements from Timing the Binary Pulsar PSR B1913+16}, by Weisberg and Huang 2016 (ApJ, Vol. 829, Issue 1, article id. 55, published in September 2016, 10 pp.) -- $\textcopyright$ AAS. Reproduced with permission.}

\subsection{Searching for GWs in the Cosmic Microwave Background}

The cosmological inflation is an epoch in the early history of the Universe, that is conjectured to seed structure formation and primordial GWs. The quasi-exponential expansion of the Universe during this phase is thought to have generated a stochastic background of GWs \citep{sta79}, that cannot be detected directly with current instrumentation. However, indirect detections may be possible. The inflationary GW background is predicted to have excited both of the polarization patterns of the Cosmic Microwave Background (CMB); the E-mode pattern (curl-free) and the B-mode pattern (curl). Although the GW-induced E-mode it is not expected to be detectable, the signature in the otherwise quiescent B-mode should be measurable \citep{pol85} as an excess power at large angular scales (the \textit{recombination bump} at $l\sim100$, where $l$ is the multipole moment). A detection of the B-mode would provide crucial information in support of the inflationary model. However, this achievement is challenging because of B-mode contaminations given by the gravitational lensing of the E-mode on small angular scales ($l\sim1000$), and the polarized foreground emission (such as from dust and synchrotron radiation) from our Galaxy \citep{tmv05} on spatial scales that are searched for the inflationary signature.

Searches for the B-mode of the CMB polarization are currently ongoing, through experiments such as POLARBEAR \citep{kaa12}, the ongoing observations with the South Pole Telescope (SPT, see e.g. \citealt{baa14}), and the Background Imaging of Cosmic Extragalactic Polarization (BICEP, see e.g. \citealt{kab03}). These experiments, that are focusing on smaller $l$ than the satellites, are expected to detect effects of the inflationary GW background at ultra-low frequencies (below $10^{-16}$ Hz, \citealt{lms16}). No detection of the inflationary-induced B-mode has been made to date, while the E-mode was observed for the first time with the Degree Angular Scale Interferometer \citep{kl02} and followed up in more detail by the Wilkinson Microwave Anisotropy Probe (WMAP) and the Planck satellites (see e.g. \citealt{ksd11,pla14}). 
%

\subsection{Searching for GWs with interferometers}

In addition to indirect detections based on pulsars or the CMB, it is possible to make direct detections of GWs. This has been achieved by the Advanced Laser Interferometer Gravitational-wave Observatory (aLIGO), and in the future, direct detections will be possible with the Evolved Laser Interferometer Space Antenna (eLISA), and PTAs. \\

aLIGO comprises two laser interferometers (in Hanford and Livingston) capable of detecting changes in the length of the interferometer arms induced by GWs. As other ground-based laser interferometers such as Virgo in Italy \citep{aaa15}, GEO 600 \citep{waa02} in Germany or KAGRA (\citealt{atm17}, not online yet) in Japan, aLIGO explores GW frequencies from approximatively $1$ to $10^3$ Hertz. The lower limit is set by gravity gradients in the Earth gravitational potential, while the upper limit is given by the shot noise of the laser photons \citep{aaa13}. GWs emitted in this frequency range are predicted to be generated by coalescing binary systems of neutron stars or stellar-sized black holes. In September 2015, aLIGO achieved the first direct detection of GWs \citep{aaa16} from a coalescing binary of stellar-mass black holes. This detection (followed by other six events and one candidate since 2015, \citealt{aaa16b,aaa17b}) signed the beginning of the era of GW astronomy. \\
However, more is required to explore this branch of science -- more detections, and a wider range of GW frequencies. \\

eLISA \citep{eli13,asa17} is a project of the European Space Agency to deploy a $3-$body, space-based interferometer with arms $2.5$ million kilometres long, that will probe the GW spectrum in a frequency range from $10^{-1}$ down to $10^{-5}$ Hz. An eLISA pathfinder, a 40-cm one-armed miniature of the future device, was launched in $2015$ and reported significantly lower noise levels than expected \citep{aaa17}. This success grew confidence and expectations for the mission, planned for launch in 2034. The frequency range limit of eLISA is given by the measurement accuracies of the free-falling test mass accelerations \citep{asa17}. The predicted sources of GWs in this frequency range are inspiralling binary systems of white dwarfs and super-massive black holes (SMBH). \\

The only experiment that can currently provide longer interferometric baselines, on a parsec scale, are PTAs (see Section~\ref{sec:pta}). PTAs explore the frequency range from about $10^{-6}$ to $10^{-9}$ Hertz, where the most likely source of GW emission are coalescing SMBH binaries (SMBHB). Other sources might be cosmic strings and relic GWs from inflation. PTAs are experiments based on the monitoring of an ensemble of selected pulsars, in order to search for spatially-correlated deviations in the arrival times of their pulses. A number of phenomena can induce such correlations, included GWs.\\

\begin{figure*}
 \begin{center}
 \includegraphics[trim=0 0 0 3cm,scale=0.7]{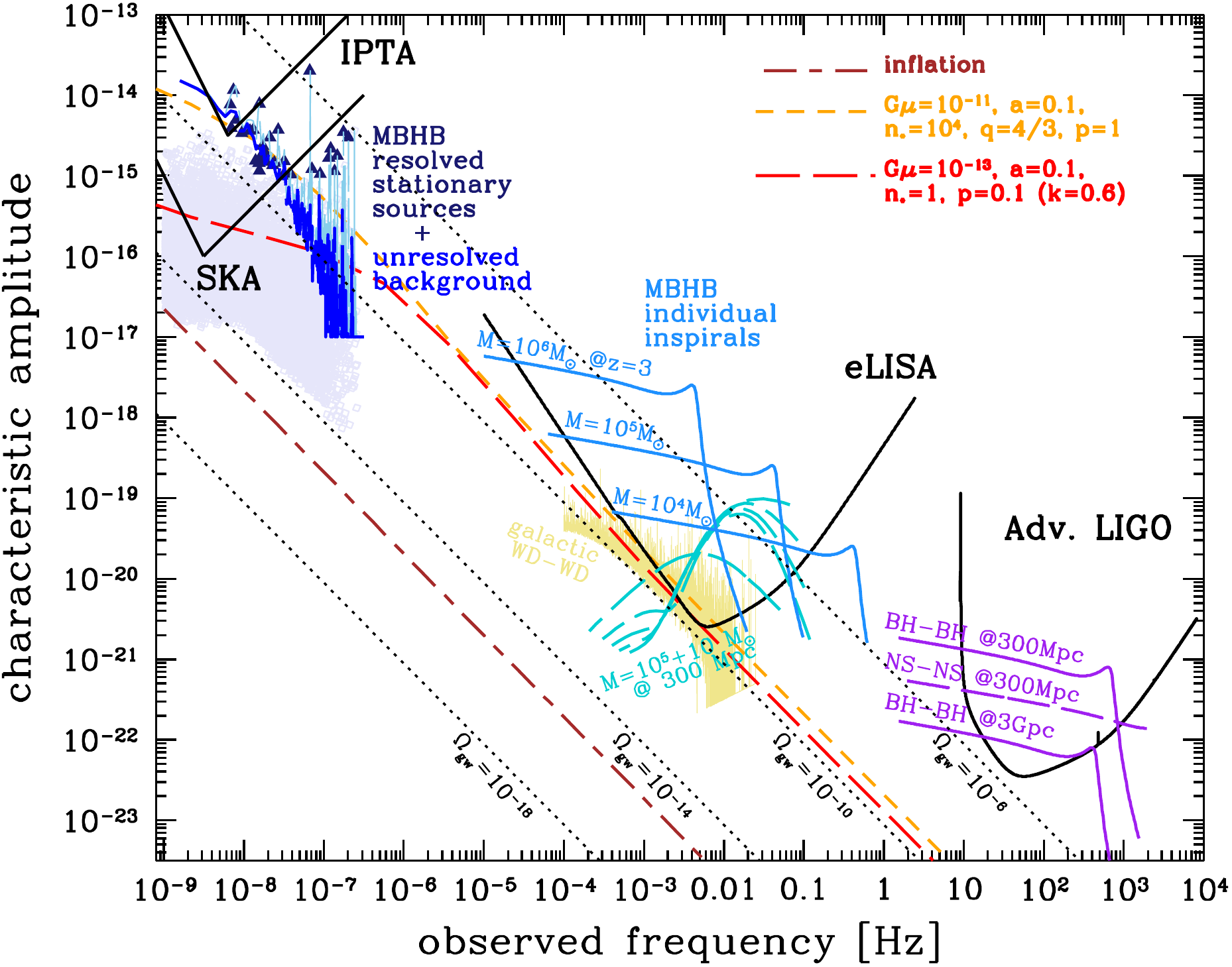}
 \caption{Figure taken from \citealt{jhm15}\protect\footnotemark; GW amplitude versus GW frequency, and frequency ranges explored by the interferometric experiments searching for GWs, aLIGO, eLISA and PTAs. In the ``PTA band'', the nominal sensitivities for the International Pulsar Timing Array are shown and the Square Kilometer Array, together with a representation of the expected emission from the SMBHB population (solid blue line) in the universe, the emission from ``GW-loud'' SMBHBs (blue triangles) and from the unresolvable SMBHBs (light purple squares). In the ``eLISA band'', the nominal eLISA sensitivity curve is shown, together with the expected GW signals from different masses of merging SMBHBs (cyan), a binary with a very high mass ratio (aquamarine), and from the Galactic population of inspiralling white dwarf binaries (yellow). In the ``aLIGO band'' the sensitivity curve of aLIGO (as of 2015) is shown, together with the expected signals from different inspiralling compact-object binaries (purple). In brown, orange and red are the GW background expected from inflation and two models of cosmic strings.}
 \label{fig:frequencycoverage}
 \end{center}
 \end{figure*}

The GW frequency bands explored by these three kinds of experiments are complementary, as shown in Figure~\ref{fig:frequencycoverage}.

\footnotetext{Figure 1 of \textit{Gravitational wave astronomy with the SKA}, by Janssen et al. 2015 (Proceedings of Advancing Astrophysics with the Square Kilometre Array (AASKA14). 9 -13 June, 2014. Giardini Naxos, Italy. Online at \url{http://pos.sissa.it/cgi-bin/reader/conf.cgi?confid=215}, id.37)}

\section{Pulsar timing}\label{sec:timing}

As mentioned in Section~\ref{sec:intro}, pulsars are often referred to as ``cosmic clocks'', as it is possible to predict their times of arrival (ToAs) to high accuracy. 
Essentially, there are three requirements to enable accurate predictions of the arrival times:

\begin{enumerate}
 \item \textit{The pulsar is a stable rotator}. As mentioned in Section~\ref{sec:intro}, pulsars lose rotational energy via magnetic dipole radiation, and therefore they spin down. However, the spin-down might suffer from irregularities, in the form of abrupt (``glitches'', \citealt{dow81}) or long-term (``timing noise'', see Section~\ref{sec:timing}) variations in the spin frequency. For high-precision experiments, pulsars should have predictable spin-evolutions;
 \item \textit{The shape of the integrated pulse profile is stable in time}. Pulsars are intrinsically weak sources, with fluxes in the order of a few milliJansky \citep{lylg95,kxc+98}. Typically, the individual pulses do not exceed the radiometer noise of the telescope. Thus, many pulsar studies use the integrated pulse profile, i.e. the coherent sum of many thousands of individual pulses. While individual pulses differ often from each other (both in flux distribution and phase), the integrated profile is statistically stable. Known sources of variations in the integrated profile are pulse jitter \citep{lb16}, plasma propagation effects \citep{gk17} or magnetospheric instabilities \citep{lhk10}. The long-term temporal stability of the integrated profile is an assumption for high-precision experiments, and efforts are ongoing to mitigate the impact of integrated profile variations in PTA experiments \citep{lk17};
 \item \textit{The timing model of the pulsar is well known}. The timing model of a pulsar (or ``ephemeris'') is a set of parameters that describes the pulsar spin and spin-down, its orbital parameters (if any), its astrometry, and the dispersive influence of the ionised interstellar medium (IISM) along the line of sight to the pulsar. The frequency-dependent dispersive effect of the IISM on radio pulses is quantified by the dispersion measure. The dispersion measure (DM) is defined as the integrated column density of free electrons along the line of sight:
 \begin{equation}\label{eq:dm}
  \mathrm{DM} = \int_{0}^{\text{d}} n_e dl
 \end{equation}
\noindent where $d$ is the distance to the pulsar (pc), and $n_e$ is the free electron number density (cm$^{-3}$).
\end{enumerate}

The first draft of ephemeris for a certain pulsar can be obtained from its discovery, and provides an approximate estimate of the pulsar spin, position and DM. A precise knowledge of the timing model can be achieved through the technique of \textit{pulsar timing} \citep{lk05}. \\
Let us assume that an observing campaign is performed on a pulsar with a given radio telescope. For each observation, we can obtain an integrated pulse profile $P$, so that the pulse profile is statistically stable and has a suitably high signal-to-noise ratio (S/N). In pulsar timing, average ToAs for each observation are computed via a cross-correlation of each integrated pulse profile with a high-S/N reference template $S$ \citep{tay92}, which is typically a noise-free representation of the pulse profile. This yields a phase shift $\tau$ between $P$ and $S$, if we consider $P$ to be described as:

\begin{equation}
 P(t) = a + bS(t - \tau) + n(t),
\end{equation}

\noindent where $a$, $b$ and $n(t)$ represent, respectively, an intensity baseline, an intensity normalization, and the noise level. The \textit{topocentric} ToA, $ToA_{\text{topo}}$ is the sum of $\tau$ to a time stamp associated with the observation. \\
The topocentric ToAs are then transformed to the (at first order) inertial reference frame of the Solar System barycentre (SSB). This conversion is based on the parameters included in the timing models, a reference for the time standard, and for the planetary ephemeris \citep{ehm06}:

\begin{equation}
ToA_{\text{SSB}} = ToA_{\text{topo}} + t_{\text{clk}} - \frac{D}{f^2} + \Delta_\text{R} + \Delta_\text{E} + \Delta_\text{S}.
\end{equation}

\begin{figure}
\begin{center}
\includegraphics[scale=0.377]{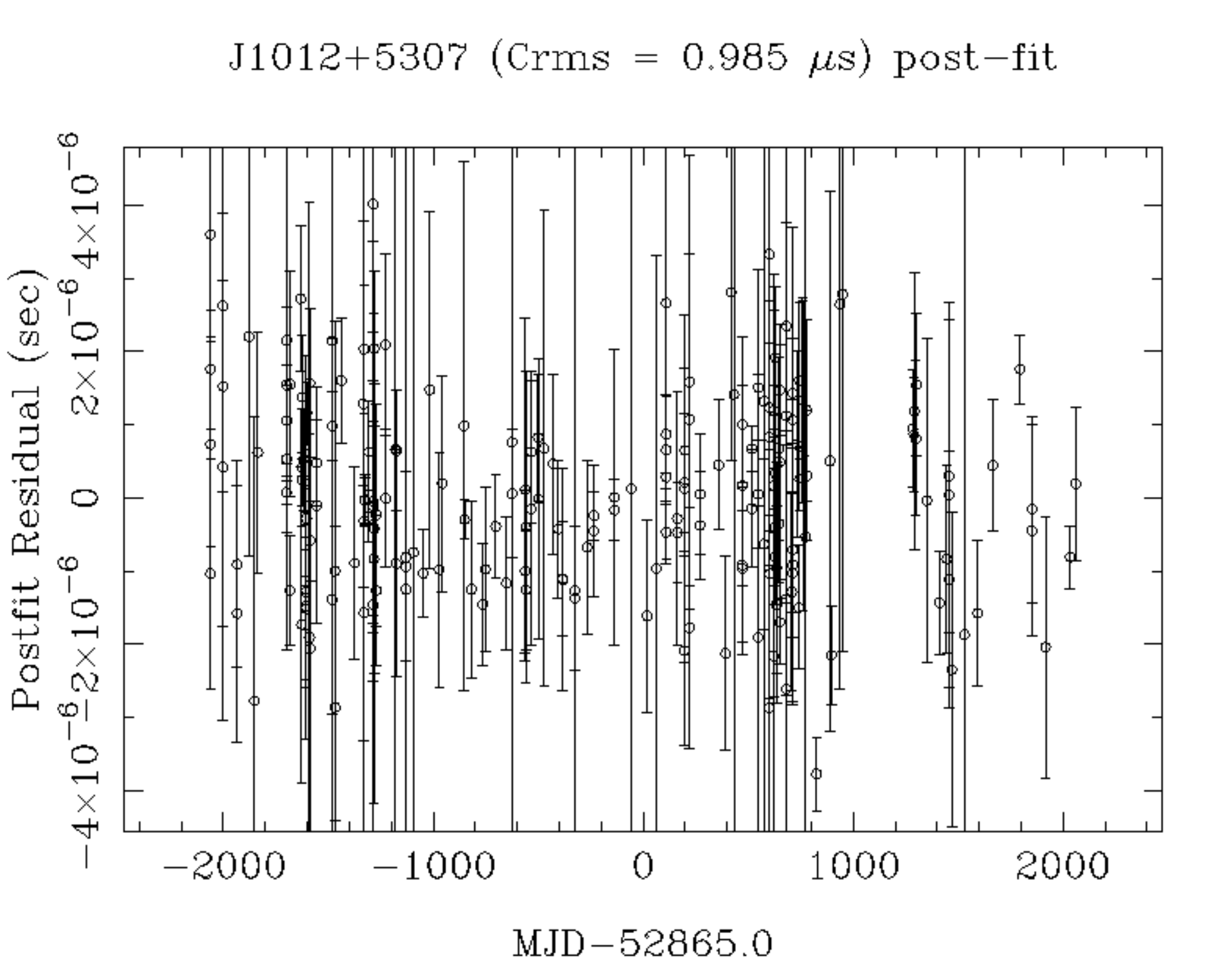}
\caption{Timing residuals versus time for PSR\,J1012+5307. The used observations were obtained at L-band with the Effelsberg radio telescope, and the Effelsberg-Berkeley Pulsar Processor (EBPP) backend. The used ephemeris, with no additional fitting applied, were obtained by \citet{vlh16}, based on all the available IPTA datasets, including the EBPP one (i.e., the IPTA data release, see Section~\ref{sec:results}).}
\label{fig:whiteresiduals}
\end{center}
\end{figure}

\begin{figure}
\begin{center}
\includegraphics[trim=-0.4cm 0 0 0,scale=0.38]{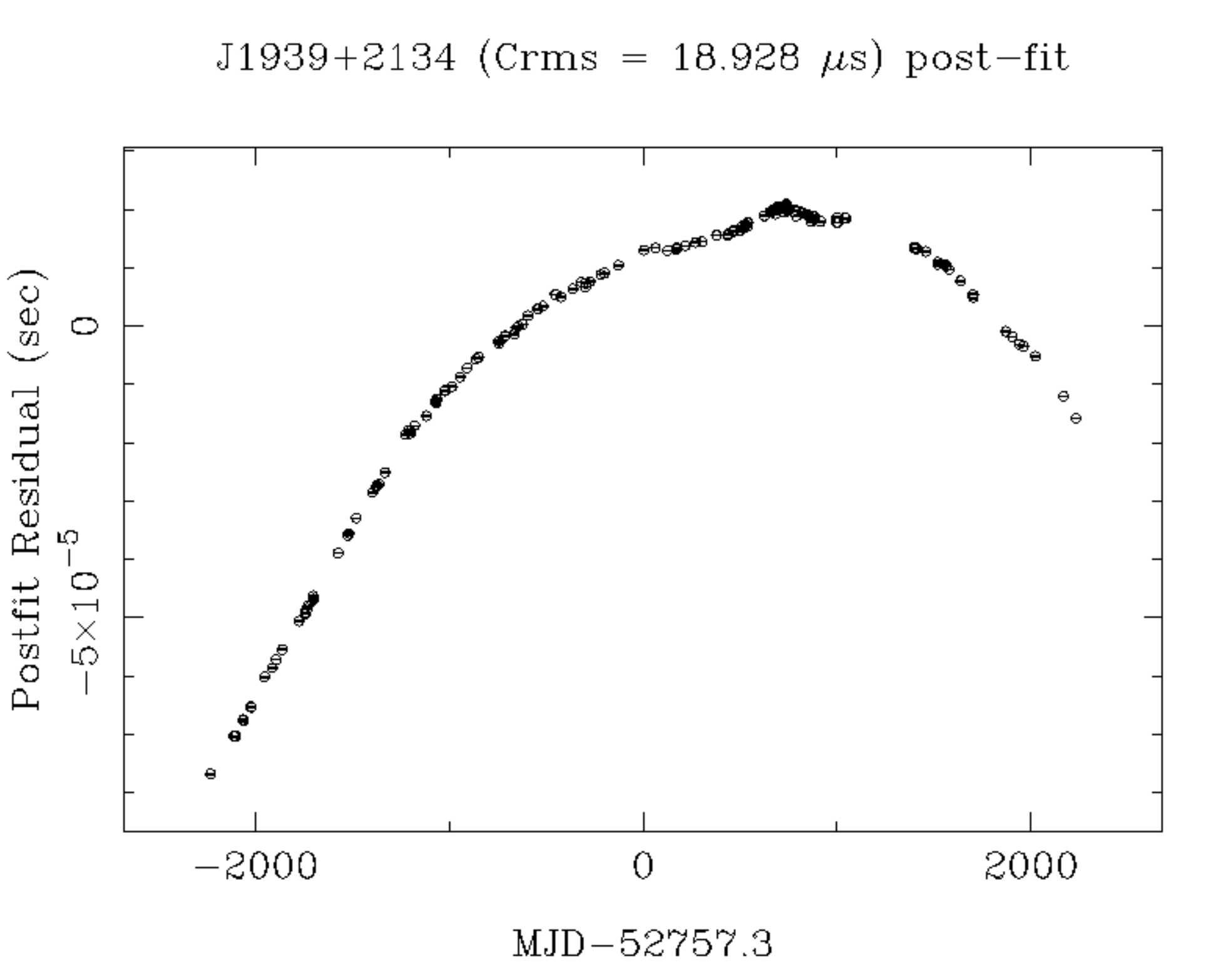}\\
\includegraphics[trim=-0.4cm 0 0 0,scale=0.255]{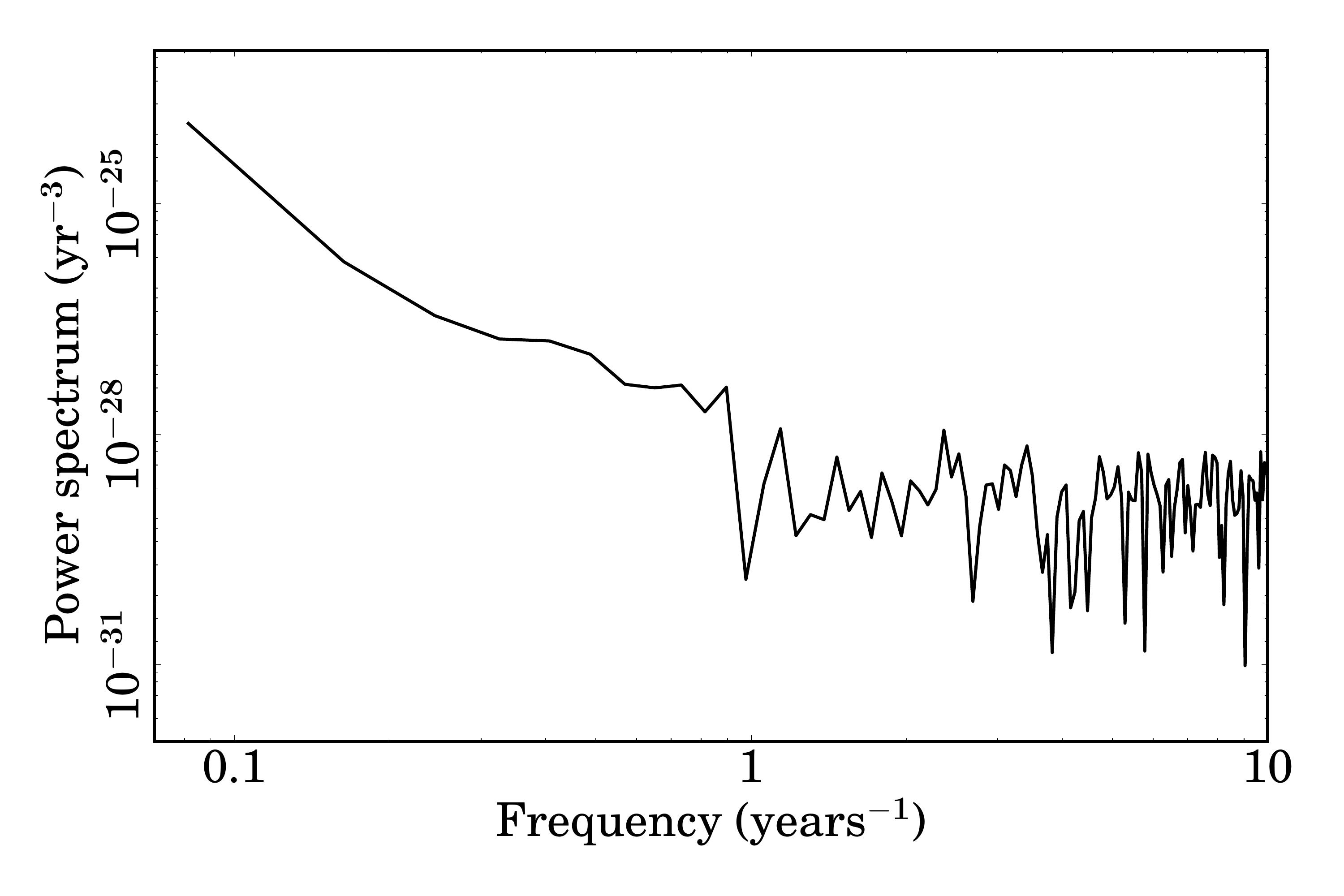}
\caption{Timing residuals versus time for PSR\,J1939+2134 (\textit{upper panel}), and corresponding power spectrum versus frequency (\textit{lower panel}). The used observations were obtained at L-band with the Effelsberg radio telescope, and the Effelsberg-Berkeley Pulsar Processor (EBPP) backend. The used ephemeris, with no additional fitting applied, were obtained by \citet{vlh16}, based on all the available IPTA datasets, including the EBPP one (i.e., the IPTA data release, see Section~\ref{sec:results}). The timing residuals of PSR\,J1939+2134 are clearly affected by red noise, most likely spin noise.} 
\label{fig:redresiduals}
\end{center}
\end{figure}

\noindent In this equation, $t_{\text{clk}}$ transforms the reference time standard from the (typically) maser-based clock at the observatory to a world-wide recognized time standard such as Terrestrial Time. The third term removes the effects of observing at non-infinite frequency:

\begin{equation}
 D = \frac{e^2}{2 \pi m_e c} DM,
\end{equation}

\noindent where $e$ and $m_e$ are the charge and the mass of an electron, and $c$ is the light speed. $\Delta_\text{R}$ is the Roemer delay, that corrects for the difference in travel time between the observatory and the SSB. The Roemer delay is purely based on geometrical considerations, and uses the astrometric parameters of the studied pulsar and the planetary ephemeris. $\Delta_\text{E}$, the Einstein delay, is based on the planetary ephemeris and corrects for the effects of the gravitational redshift induced by the bodies of the Solar system. $\Delta_\text{S}$, the Shapiro delay, accounts for the additional time travel required to the light waves for travelling across the gravitational field of the Solar system. Additional corrective parameters are required if the pulsar is part of a binary system.\\
Once the barycentric ToAs $t$ has been derived, we compute the pulse number $N$ that represents a ``counter'' for the number of pulsar rotations:

\begin{equation}\label{eq:rotationcounter}
 N (t) = N_0 + \nu_0 (t - t_0) + \frac{1}{2}\dot{\nu}(t - t_0)^2 + ...
\end{equation}

\noindent where $N_0$ is the pulse number at the reference time $t_0$, and $\nu_0$ and $\dot{\nu}$ are the spin period at $t_0$ and the spin down rate respectively. The right hand side of Equation~\ref{eq:rotationcounter} is the Taylor expansion of the pulsar spin. \\
In the last steps of a timing analysis, the parameters included in the timing model can be varied so that the ToAs are spaced of an integer number of pulsar rotations. The refinement of the timing model can be achieved through dedicated software such as \textsc{tempo2} \citep{hem06} and the inspection of the ``timing residuals'', i.e. the difference between the closest integer number of pulsar rotations and actual number of pulsar rotation among the ToAs. If parameters in the timing model are imprecisely estimated or missing, then we expect to see structures in the timing residuals. For example, we see from Equation~\ref{eq:rotationcounter} that an incorrect spin frequency or spin-down rate will show as a linear and a parabolic trend in the timing residuals (see \citealt{lk05}). If the timing model is sufficiently accurate, then the timing residuals will look ``white'', i.e. with no correlations (see Figure~\ref{fig:whiteresiduals}). For reasons that will be explained in Section~\ref{sec:pta}, PTAs are particularly interested in the study of the ``red noise''. A time series affected by red noise shows long-term correlated structures in the time domain, and an excess in the low frequency bins of its power-spectrum (see Figure~\ref{fig:redresiduals}). Several phenomena can induce red noise, for example DM variations \citep{yhc+07}, or intrinsic instabilities in the pulsar spin (better known as ``spin noise'' or simply ``timing noise'', \citealt{cll16}), instrumental imperfections, or gravitational waves. \\

\section{Sources of gravitational waves at low frequencies}\label{sec:sources}

As mentioned in Section~\ref{sec:intro}, GWs are produced by the second time derivative of the quadrupole moment of the mass distribution that distorts the space-time.  \\
Following a dimensional analysis, the amplitude $h$ of a GW is given by \citep{hug03}:

\begin{equation}
 h \propto \frac{G}{r c^4} \frac{d^2 Q}{dt^2}
\end{equation}

\noindent Where $G$ is the gravitational constant, $r$ is the distance to the GW source and $Q$ is the quadrupole moment. Due to the factor $G/c^4$, the peak GW amplitude is expected to be small. GWs are therefore more easily detectable when the second derivative of the quadrupole moment is large \citep{tho87}, as in the case of massive, fast-moving objects. \\
The most likely source of GWs at low frequencies are coalescing SMBHBs, although other potential sources have been identified: GWs from inflation \citep{gri74,sta79} or cosmic strings \citep{kib76}.

\subsection{Super-massive black hole binaries}\label{sec:sources_smbhb}
Observational evidence shows that SMBHs are hosted at the centre of the most or all galaxies \citep{kr95,mtr98}. The hierarchical or ``bottom-up'' scenario \citep{wr78} predicts that larger galaxies are generated via merging of smaller galaxies at high redshifts ($z$). When two galaxies merge, we then expect that the two SMBHs at their centres form a binary system \citep{bbr80,vhm03}. A binary system is characterized by a non-zero value of the second time derivative of its mass quadrupole moment, thus it is  a continuous source of GWs. Defining the masses of the individual SMBHs as $m_1$ and $m_2$, and assuming a circular orbit for simplicity, with total mass $M = m_1 + m_2$, reduced mass $\mu = m_1m_2/M$ and $\mathcal{M}=(m_1m_2)^{3/5}/M^{1/5}$ (the \textit{chirp mass}), and using geometricised units such that $G=c=1$, the luminosity emitted in GWs ($L_{\text{gw}}$) by SMBHB is given by \citep{tho87,ses13a}:

\begin{equation}\label{eq:smbhbgwluminosity}
 L_{\text{gw}} = \frac{32}{5} (\pi \mathcal{M} f)^{10/3}
\end{equation}

\noindent where $f$ is the observed frequency of the emitted GWs, equal to twice the orbital frequency $f_\text{B}$. 

The inclination-polarization averaged amplitude, $h$, of the radiated GWs is given by \citep{ses13a}:

\begin{equation}\label{eq:smbhbgwamplitude}
 h = \sqrt{\frac{32}{5}} \frac{G \mathcal{M}}{r c^4} (\pi f)^{2/3}
\end{equation}

Equations~\ref{eq:smbhbgwluminosity} and \ref{eq:smbhbgwamplitude} describe the simple case of a SMBHB in the local universe (i.e. with zero redshift). As pointed out by \citet{vec04}, it is possible to ``move'' the GW source at a different redshift by substituting $m_x$ with $m_x(1+z)$, $r$ with $r(1+z)$ and $f$ with $f(1+z)$. \\
Note that both of the expressions for the GW luminosity and strain contain the GW frequency $f$. 
During the binary inspiral, $f_\text{B}$ (and hence $f$) changes in time. This means that the strain $h$ of the propagated GW might not be the same in two different points in space time. However, during short time scales over which the change in orbital separation is negligible, the SMBHB can be considered a monochromatic source of GWs \citep{sv10}.\\
We can identify three stages in the evolution of a SMBHB \citep{fh98}: 

\begin{enumerate}
 \item \textit{Inspiral}, The two SMBHs orbit each other and such orbital separation shrinks due to environmental effects and GW emission;
 \item \textit{Merger}, The two SMBHs coalesce, emitting a GW burst that permanently modifies space-time, called a ``memory'' event;
 \item \textit{Ringdown}, the SMBH and the nearby space-time undergo to a relaxation that leads to a spherical configuration.
\end{enumerate}

\noindent $f_{\text{gw}} = 2f_B$, the low-frequency GW emission is supposed to happen during the inspiral and merger phases, while the ringdown stage is predicted to generate GWs at higher frequencies.\\

The ``memory'' phenomenon \citep{bt87} is a non-oscillatory GW emission that should occur before the actual coalescence of the two BHs. The final stages of the merger generate a net non-zero contribution to the ``plus'' polarization mode of the GW emission. Such a DC offset induces a permanent deformation in the metric \citep{fav09}. Memory events can only be detected during their passage through the detectors, that, in the moments when they operate the metric deformation. The strain of a memory event, $h_{m}$, is predicted to be \citep{mcc14}:

\begin{equation}
 h_m \approx \frac{1-\sqrt{8}/3}{24} \frac{G\mu}{c^2 r} \sin{\mathcal{I}}(17 + \cos^2{\mathcal{I}})\times[1 + \mathcal{O}(\mu^2/M^2)]
\end{equation}

\noindent where $\mathcal{I}$ is the inclination angle of the binary before the merger. For $r=1$ Gpc, and $M_1=M_2=10^9$ solar masses, $h_m \approx 10^{-15}$ \citep{mcc14}.

The amplitude of a memory event should rapidly increase in the very final stages of the coalescence before the merging, in a timescale $\tau \approx 2 \pi R_s/c$, where $R_s$ is the Schwarzschild radius \citep{cj12,mcc14}.

\section{Pulsar Timing Arrays}\label{sec:pta}
Here we outline the expected signatures in pulsar timing data given by the various types of GW emission from SMBHBs.
\subsection{Modelling a GW signal from single sources}
Two processes are predicted to lead to the orbital shrinking of the SMBHB. First, the shrinking is led by environmental effects until the GW emission reaches approximatively nHz frequencies (see \citealt{ses13a} for a review). Let us assume a circular, evolving (i.e. shrinking) SMBHB, and a pulsar $p$, characterized by an angle $i$ between the orbital plane of the SMBHB and the line-of-sight toward the pulsar. Because the GW emission from the SMBHB permeates the entire sky in a quadrupolar fashion, we can expect that ToAs from $p$ will be affected, arriving slightly advanced or slightly delayed than the timing model prediction \citep{saz78}. It can be demonstrated that the effect on the timing residuals $R(t)$ of a single pulsar is independent from the travel path of the radiation \citep{det79}. In particular \citep{bps15}:

\begin{equation}\label{eq:gwtimingresponse}
 R(t) = R_\text{E}(t) - R_\text{p}(t),
\end{equation}

\noindent $R_\text{E}(t)$ and $R_\text{p}(t)$, called the \textit{Earth term} and the \textit{pulsar term}, describe the residuals induced by GWs passing over the Earth and the pulsar respectively. In the quadrupolar approximation \citep{lom15} we have that \citep{bps15}:

\begin{multline}
 R_\text{E}(t) = \frac{h}{\omega} \{(1+cos^2{i})F^+[sin(\omega t + \Phi) - sin{\Phi}] \\ + 2cos{i}F^{\times} [cos(\omega t + \Phi) -cos{\Phi}] \}, \\
 R_\text{p}(t) = \frac{h_\text{p}}{\omega_\text{p}} \{(1+cos^2{i})F^+[sin(\omega t + \Phi + \Phi_\text{p}) - sin(\Phi \\ + \Phi_\text{p})] + 2cos{i}F^{\times} [cos(\omega t + \Phi + \Phi_\text{p}) -cos{\Phi + \Phi_\text{p}}] \},
\end{multline}

\noindent where $h$ and $h_\text{p}$ are the amplitudes of the GW at the Earth and at the pulsar, $\omega$ and $\omega_\text{p}$ are the GW angular frequency at the Earth and at the pulsar, $\Phi$ and $\Phi_\text{p}$ are the GW phases at the Earth and at the pulsar, and $F^+$ and $F^\times$ are the antenna response functions for the two GW polarization at the pulsar (i.e. how space-time around the pulsar is affected by the GW). If the SMBHB does not evolve, the power spectrum of the signature described in Equation~\ref{eq:gwtimingresponse} is described by two Dirac delta functions. This signal would be indistinguishable from the signature given by an error in the orbital period and thus not detectable with pulsar timing.\\  

As mentioned in Section~\ref{sec:sources_smbhb}, the last stage of a SMBHB merger lead to a permanent deformation of the metric, similar to a DC offset in space-time. This non-oscillatory phenomenon is propagated, and affects the timing residuals of a pulsar in a way that is equivalent to a change in its rotational spin frequency \citep{vhl10}. The timing residuals induced by a memory event will be given by \citep{vhl10}:

\begin{equation}\label{eq:timingresidualsburst}
 R(t) = h_m B(\theta,\phi)\times [\Theta(t-t_0) - \Theta(t-t_1)],
\end{equation}

\noindent where $B(\theta,\phi)=1/2 \cos{(2\phi)}(1-\cos{\theta})$, $\theta$ is the angular separation between the pulsar and the SMBHB, $\phi$ is the angular separation between the principal polarization of the GW signal and the projected line-of-sight to the pulsar onto the plane perpendicular to the GW propagation direction. $\Theta$ is the Heaviside function, while $t_0$ and $t_1$ are the instants in which the memory event passes the Earth and the pulsar respectively. In Equation~\ref{eq:timingresidualsburst}, it is thus possible to identify an Earth term and a pulsar term, as for the oscillatory contribution to the GW emission shown in Equation~\ref{eq:gwtimingresponse}. In particular, the Earth term sensitivity to a memory event is found to increase with the square root of the number of pulsars included in a PTA \citep{cj12}.

\subsection{Modelling a GW background signal}

The expected number of SMBHB systems is extremely large, up to $10^6$ depending on the redshift and the mass range of the involved BHs \citep{svc08}. The choral GW signal coming from such a population of SMBHBs gives rise to an incoherent superposition of the individual GW signals, that effectively generates a \textit{stochastic background} of GWs (GWB \citealt{svc08,rwsh15}), usually considered isotropic.\\ 
The GWB is predicted to induce a red-noise signal in pulsar timing residuals, with a power spectrum $P(f)$ that can be described by a steep power-law \citep{phi01}:

\begin{equation}\label{eq:ps}
 P(f) = \frac{h^2}{12\pi^2} \left ( \frac{f}{f_{\text{1yr}}} \right ) ^{2\alpha -3},
\end{equation}

\noindent where $f_{\text{1yr}}$ normalizes the GW frequency at $1/1_{\text{yr}}$, $h$ is now the amplitude of the GWB, and $\alpha$ is a coefficient whose value is $2/3$ in the case of an isotropic and stochastic GWB, thus the spectral index for a GWB is expected to be $-13/3$. In the case of a GWB, $\Omega_{\text{gw}}(f)$, the ratio between the energy density $\rho_{\text{gw}}$ of the GWs (per unit logarithmic frequency) and the critical energy density of the Universe $\rho_c$, is related to the strain $h$ as \citep{ar99}:

\begin{equation}\label{eq:energydensitygwb}
 \Omega_{\text{gw}} (f) = \frac{2 \pi^2}{3 H_0^2} f^{2} h^2(f),
\end{equation}

\noindent where $H_0$ is is the Hubble expansion rate ($100\,h_H$ km s$^{-1}$ Mpc$^{-1}$, with $h_H$ being the dimensionless Hubble parameter). In the case of relic GW from inflation and cosmic strings, the spectral index of Equation~\ref{eq:ps} is expected to be $-5$ \citep{gri05} and $-16/3$ \citep{dv05} respectively.\\



\subsection{The Hellings \& Downs curve}
Although SMBHBs are considered the loudest sources of GWs in the universe, the amplitude of such emission is predicted to be extremely tiny \citep{abb14,bps15,zwx16}. The GWB amplitude is expected to exceed the amplitude of the signals from the vast majority of individual SMBHBs. This implies that the detection of a GWB is much more likely than GWs from an individual SMBHB \citep{rsg15}.\\
As already discussed, the GWB is a stochastic signal that can be described as a red noise process (Equation~\ref{eq:ps}).  Given an individual pulsar, a GWB signal cannot be distinguished unequivocally from other red noise processes such as timing noise, IISM effects, clock noise, or ephemeris errors. Additionally, the pulsar timing procedure absorbs all of the power present in the first two bins of the power spectrum of the timing residuals (corresponding to the spin period and spin period derivative) and in the 1/1yr frequency bin (due to the orbital period of the Earth), effectively decreasing the detectability of a GW signal at this frequency. \\

However, a GW emission (both from SMBHBs and a background) would affect different pulsars in a correlated fashion depending on their respective sky position. Therefore, detection statistics are based on the correlation among the timing residuals of an \textit{array} of pulsars \citep{rom89,fb90}. \\

\begin{figure}
\begin{center}
\includegraphics[scale=0.45]{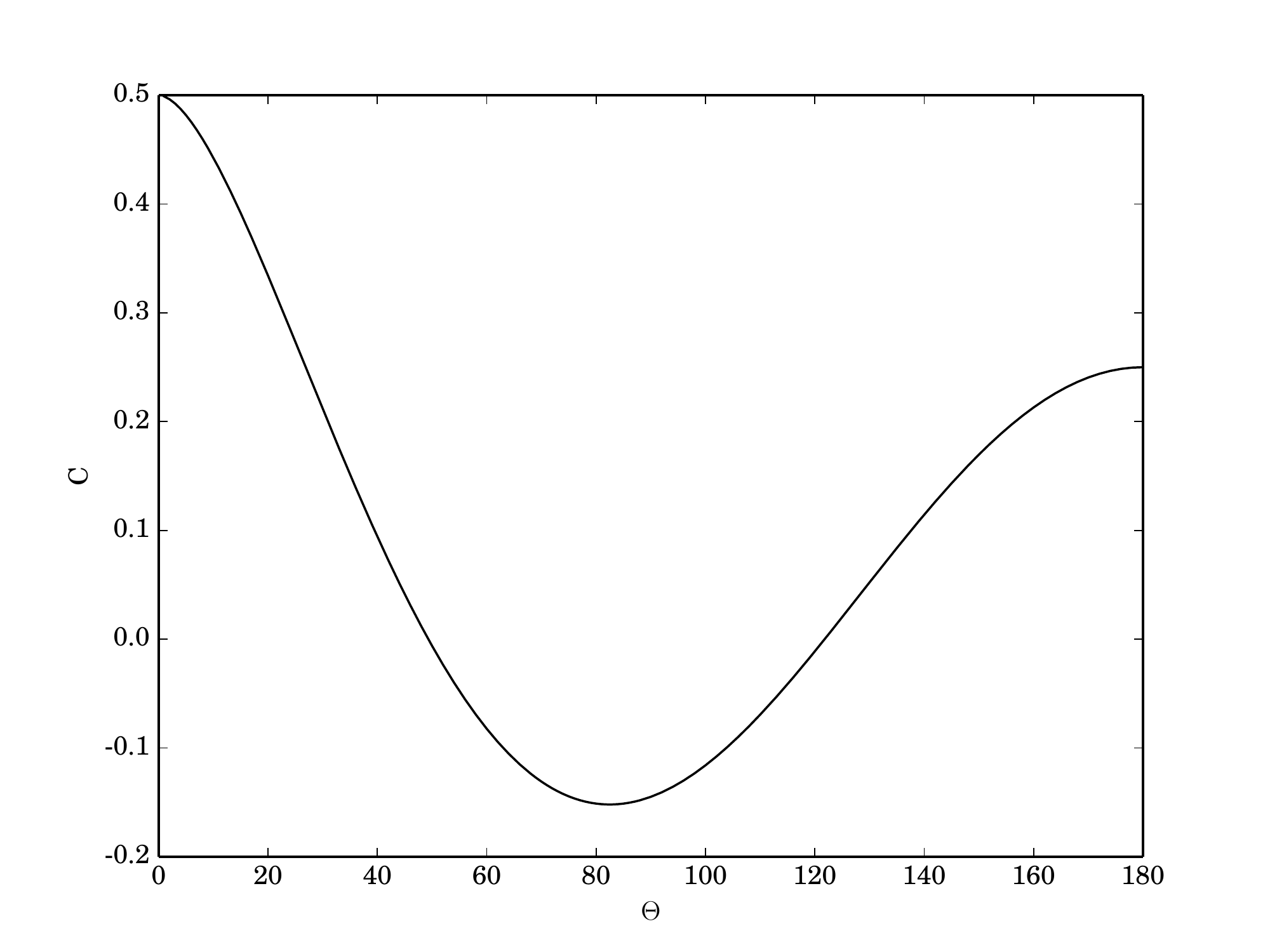}
\caption{Angular correlation $C$ given by the Hellings \& Downs curve as described by Equation~\ref{eq:hdcurve} (minus the contribution of the pulsar term) versus angular distance $\theta$.}
\label{fig:hd}
\end{center}
\end{figure}

The correlation $C$ among the timing residuals of pairs of pulsars perturbed by an isotropic and stochastic GWB was studied by \citet{hd83}. Given a pair of pulsars $(i,j)$, separated by an angle $\theta_{\text{i,j}}$  in the sky, they demonstrated that $C$ takes a specific functional form, known as ``Hellings \& Downs curve'':

\begin{equation}\label{eq:hdcurve}
 C(\theta_{ij}) = \left[ \frac{3}{2} x\, \text{log}(x) - \frac{x}{4} + \frac{1}{2} \right](1+\delta_{\text{i,j}}),
\end{equation}

\noindent where $x = (1 - cos{\theta_{\text{i,j}}})/2$. The Hellings \& Downs curve is the sky- and polarization-averaged angular correlation between pairs of pulsars. In the computation, the GWB is assumed to be isotropic (i.e. the power spectrum of the GWB does not have an angular dependence), and the short-wavelength approximation to be valid (i.e. $f_{\text{gw}}r>>1$, that is, the distance between the Earth and the pulsar, and between the pulsars in the array, is large if compared to the wavelength of the GWs). It should be noted that the Hellings \& Downs curve is computed using the Earth term only. The pulsar term is estimated to bring a significant contribution only at angular distances close to zero, and only if the pulsar pair is effectively close in space. In this case, the contribution of the pulsar term brings the angular correlation to $1$. For pulsar pairs at even smaller angular distances, the contribution of the pulsar term becomes rapidly negligible \citep{ms14}. Figure~\ref{fig:hd} shows the Hellings \& Downs curve, without taking into account the additional correlation that would occur at $\theta_{\text{i,j}}=0$ when considering the pulsar term.  

\subsection{Aims and characteristics of Pulsar Timing Arrays}

PTA experiments aim to detect signals that are angularly-correlated across the sky, using the clock-like behaviour of an array of hyper-stable pulsars. The primary goal of PTAs is the detection of low frequency GWs, and the most likely GW source to emit in the PTA band is coalescing SMBHBs.
In this sense, PTAs can be considered as interferometer on Galactic scales, although instead of lasers, PTAs exploit the pulsed radio emission from the pulsars in the array. \\
To aid our detection prospects, we select pulsars with high rotational-stability for PTA analysis \citep{sc10}. The most rotationally-stable pulsars are millisecond pulsars (MSPs; \citealt{acrs82}). MSPs are pulsars that have been spun up via a transfer of mass and angular momentum by a companion star, which accelerates the neutron star to spin periods in the order of milliseconds \citep{bv91}. Following the mass transfer, both the magnetic field intensity and the spin-down rate are remarkably lowered. 
Millisecond pulsars are much more stable \citep{vbc09} than normal pulsars, and are characterized by timing residuals that are typically lower and whiter \citep{hlk+04}. As such, they are the only class of pulsars that are included in PTA monitoring campaigns.\\

\begin{figure}
\begin{center}
\includegraphics[scale=0.41]{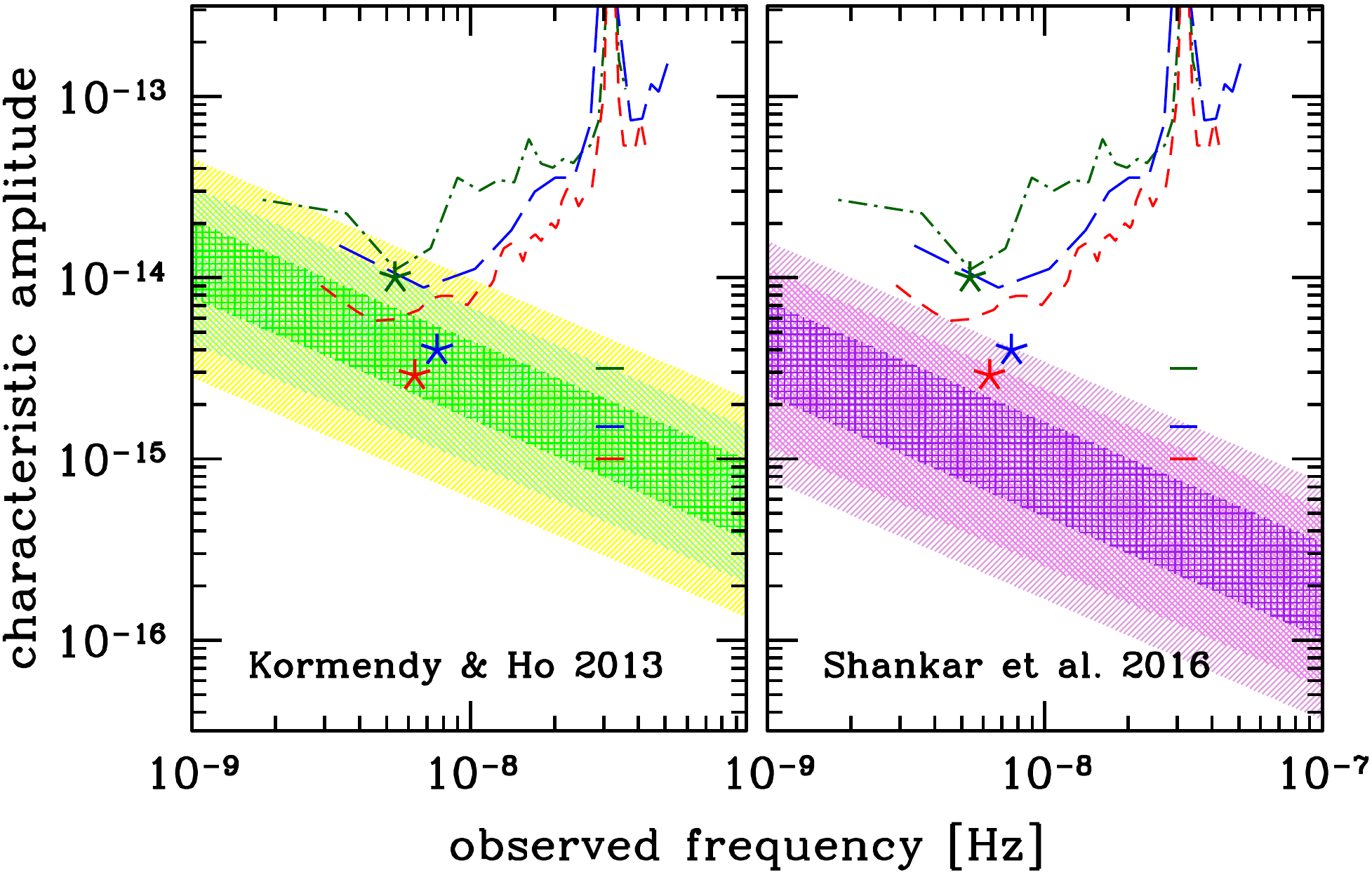}
\caption{Figure taken from \citet{ssb16}\protect\footnotemark; GWB amplitude versus $f_{\text{gw}}$. The plot shows a comparison between the GWB predictions as based on the BH-host relations from \citet{kh13} (\textit{left panel}) and \citet{sbs16} (\textit{right panel}). The sensitivity curves for EPTA, PPTA and NANOGrav are shown in green, blue and red respectively, and the differently shaded area represents $99.7\%$, $95\%$ and $65\%$ of probability.}
\label{fig:bh-hostrelation}
\end{center}
\end{figure}

The sensitivity of PTA experiments lies in the frequency range from approximatively $10^{-6}$ to $10^{-9}$ Hertz. The two boundaries are due to the limits imposed by the Nyquist theorem, and are set by the observing cadence at the higher frequency (assumed to be once per month) and the total timespan at the lower frequency (assumed to be around 20 years.\\
No GW detection has yet been made by PTA experiments. However, the upper limits on the GW amplitudes estimated by PTAs have already given powerful insights in the models for Galaxy formation, aiding to exclude a fraction of them \citep{ses13b,srl15}.\\
With the current sensitivities, and amplitude and rate predictions, it is unlikely that PTAs will detect \textit{individual} SMBHBs in the near future, either in the form of continuous wave or in the form of a memory event \citep{bps15,rwsh15,whc15}. Concerning a GWB, as mentioned, although its amplitude is thought to be higher than that of individual SMBHB, the peak strain is expected to be very low. \citet{ssb16}, an update of \citet{ses13b}, estimated the spectrum of the GWB amplitude generated by a population of GW-driven, adiabatically inspiralling SMBHBs in quasi-circular orbits, and demonstrated that the majority of the GWB is due to major mergers, where the mass ratio between the two SMBHs is $>0.25$ within $z=1.5$, and for black hole masses larger than $10^8$ M$_{\odot}$ \citep{svc08}. The study also assumes values from different studies available in literature to account for the SMBH binary merger rates and masses. Combining the values from observational constraints, the authors generated more than 2500 realizations of a GWB and computed a distribution for its amplitude. At $3\sigma$, they predict $1.4\times10^{-16} < A < 1.1\times10^{-15}$ at $95\%$ confidence. Such uncertainty mainly stems from poorly-constrained estimates for the galaxy merger rate and the relation between the mass of the SMBH and the mass of the host galaxy. The influence of the SMBH-host relations is shown in \citet{ssb16}, where the authors compare the GWB predictions obtained by using two of these relations (\citealt{kh13} and \citealt{sbs16}). \citet{kh13} is claimed to be biased high, due to a number of overestimated SMBH masses obtained through dynamic measurements, while \citet{sbs16} claims to have corrected the bias. The results, shown in  Figure~\ref{fig:bh-hostrelation}, indicate that the GWB predictions based on the two BH-host relations differ by a factor 3. This highlights the importance of the relations in these studies, and the necessity of refining them.\\

\footnotetext{Figure 2 of \textit{Selection bias in dynamically measured supermassive black hole samples: consequences for pulsar timing arrays}, by Sesana et al. 2016 (Monthly Notices of the Royal Astronomical Society: Letters, Volume 463, Issue 1, p.L6-L11) -- reused by permission of Oxford University Press. This figure is not covered by the Open-Access licence of this publication. For permissions contact \url{Journals.permissions@OUP.com}}

GWs, both in the form of emission from single sources and a background, are not the only signals that can be angularly-correlated among pulsars in a PTA. For example, imperfections in the reference time standards and in the planetary ephemeris used to identify the Solar System barycentre would induce, respectively, a monopolar and a dipolar angular correlation in the timing residuals \citep{fb90}, and the creation of a pulsar-based time reference and the improvement of planetary ephemerides are ongoing projects within the framework of PTAs \citep{hcm12,chm10}. \citet{thk16} studied the impact of correlated noise processes other than GWs (such as errors in time standards, planetary ephemeris, and unmodelled effects of instrumentation and the Solar wind) on PTA sensitivity to the GWB. The study demonstrated that, without including mitigation techniques in the detection pipelines, such signals can induce false detections (see also \citealt{tlb17}). Mitigation is feasible, especially for the monopolar signal, while the dipolar signal is more difficult to subtract without compromising the GWB search. 

\subsection{Current PTA collaborations}\label{sec:ptacollaborations}
There are currently three well-established collaborations in the world that are leading PTA experiments: the European Pulsar Timing Array (EPTA, \citealt{dcl16}) in Europe, the Parkes Pulsar Timing Array (PPTA, \citealt{rhc16,mhb13}) in Australia and the North American NanoHertz Observatory for Gravitational waves (NANOGrav, \citealt{abb15}) in the North America. EPTA, PPTA and NANOGrav, all based on MSP observations with 100-m class radio telescopes, collaborate as the International Pulsar Timing Array (IPTA, \citealt{vlh16}). 

\subsubsection{EPTA}

The EPTA was officially established in $2005$, and currently monitors $42$ MSPs \citep{dcl16} at an approximatively monthly cadence with each of the five largest radio telescopes in Europe: the Effelsberg Radio Telescope (Eff, Germany), the Nan\c{c}ay Radio Telescope (NRT, France), the Westerbork Synthesis Radio Telescope (WSRT, the Netherlands), the Lovell Telescope at Jodrell Bank Observatory (JBO, UK), and the Sardinia Radio Telescope (SRT, Italy). In addition, a special program within the EPTA, the Large European Array for Pulsars (LEAP), effectively acts as a sixth EPTA telescope \citep{bjk16}.\\
Besides the dataset collected since its establishment, the EPTA uses archival data of MSPs dating back to the 1990s, and collected under different timing proposals with ``historical'' backends and receivers (i.e. pulsar instruments now decommissioned).\\
The current observing setup of the EPTA telescopes is as follows:
\begin{itemize}
 \item \textit{EFF}. Performs coherently dedispersed observations of MSPs at three different frequencies with the PSRIX backend \citep{lkg16}: $1360$\,MHz, $2639$\,MHz and $4800$\,MHz; 
 \item \textit{JBO}. Observes MSPs with two backends in parallel, the DFB (incoherent dedispersion, see \citealt{mhb13}) and the ROACH (coherent dedispersion, see \citealt{bjk16}), at $1532$\,MHz;
 \item \textit{NRT}. Performs coherently dedispersed observations of MSPs in two frequency ranges, between $1100$ and $1800$\,MHz, and between $1700$ and $3500$,MHz with the NUPPI backend \citep{ldc14}; 
 \item \textit{WSRT}. WSRT is currently unavailable for EPTA observations, as a new backend and frontend (ARTS and APERTIF) are being commissioned. The previous setup performed coherently dedispersed observations at $345$, $1380$ and $2273$\,MHz with the PuMa II backend \citep{ksv08}. The receiver at $345$\,MHz has been officially decommissioned; 
 \item \textit{SRT}. The first official EPTA observing run commenced in $2016$, performing observations between $305$ and $410$\,MHz and between $1300$ and $1800$\,MHz (sometimes simultaneously) with a DFB and a ROACH backend;
 \item \textit{LEAP}: Performs coherently dedispersed, interferometric observations of MSPs with the five EPTA telescopes at $1396$\,MHz, using ROACH backends, and collects dual-polarization baseband data, that are then correlated offline.
\end{itemize}


\subsubsection{PPTA}
The PPTA project commenced in 2005, and currently monitors $24$ MSPs \citep{rhc16} with the \textit{Parkes Radio Telescope} (NSW, Australia) every two to three weeks. In addition to observations obtained within the PTA program, the PPTA uses data sets collected since the 1990s for other timing campaigns. Currently, the observations are carried out at three different frequencies: $3000$, $1500$ and $600$\,MHz, using a DFB (incoherent dedispersion) and the CASPSR\footnote{http://www.astronomy.swin.edu.au/pulsar/?topic=caspsr} (coherent) backends.

\subsubsection{NANOGrav}
NANOGrav was officially established in $2007$, and currently monitors $59$ MSPs (a selection that has been expanded after \citealt{abb16}) with the Arecibo Observatory (AO, Puerto Rico), the Green Bank Telescope (GBT, West Virginia, USA) and the Very Large Array (VLA, New Mexico, USA), every three or four weeks (Arecibo and Green Bank are also carrying out weekly observations of a subset of the monitored MSPs).\\
The current observing setup of the NANOGrav telescopes is as follows:
\begin{itemize}
 \item \textit{AO}. Performs coherently dedispersed observations of MSPs at $430$, $1410$ and $2030$\,MHz with the PUPPI backend \citep{dpr08};
 \item \textit{GBT}. Performs coherently dedispersed observations of MSPs at $820$ and $1500$\,MHz with the GUPPI backend \citep{dpr08};
 \item \textit{VLA}. The newest addition to the NANOGrav program, it observes MSPs between $1000$ and $2000$\,MHz and between $2000$ and $4000$\,MHz since $2017$.
\end{itemize}

\subsubsection{New PTA collaborations}
Efforts to establish PTA experiment are ongoing in India, China, and South Africa. \\

The \textit{Indian PTA} 
observes MSPs with the Ooty Radio Telescope (ORT) and the Giant Metrewave Radio Telescope (GMRT, both conventional and upgraded). In particular, conventional GMRT (that is timing $9$ MSPs together with ORT) has $32$\,MHz of bandwidth available in coherent dedispersion, while updated GMRT (that is timing 18 MSPs) has 200\,MHz of bandwidth available in incoherent dedispersion (coherently dedispersion will be available in the near future). ORT has a central frequency of $334$\,MHz and can observe with coherent dedispersion within $16$\,MHz of bandwidth (M. Bagchi, private communication). \\

The \textit{Chinese PTA} had an inaugural meeting in May $2017$. The Chinese PTA operates several $100$-m class telescopes (e.g. NSRT, Kunming, Tianma), but the two most important facilities will be the Five hundred meter Aperture Spherical Telescope (FAST, \citealt{pns01}) and the QiTai Radio Telescope (QTT). Once combined, FAST and QTT will be sensitive to a GWB amplitude of $2\times10^{-16}$ within a few years of observations \citep{lee16}.\\

\textit{MeerTIME} is an approved proposal dedicated to pulsar timing, that will use the MeerKAT telescope (South Africa, \citealt{bdb09}). MeerKAT is one of the numerous pathfinders for the Square Kilometer Array (see Section~\ref{sec:future}), and is currently under deployment. Among the planned pulsars that will be observed with an increasingly larger number of antennas, are several PTA-relevant sources. 

\section{Recent results from PTAs}\label{sec:results}

\subsection{EPTA}
The current EPTA data set is comprised of $42$ MSPs, and is presented in \citet{dcl16}. It includes updated timing solutions and ToAs spanning more than $15$ years of data for many of the presented MSPs, besides deepening the astrometric properties of the sources. \citet{cll16} studied the red noise properties of the EPTA dataset, and found a significant level of red noise in 25 MSPs. Errors in the time standards were estimated to affect for at most $1\%$ of the total noise budget, reducing the sensitivity to the GWB and resolvable SMBHBs.\\ 
 The six most stable EPTA MSPs were used to derive the upper limits on the GWB amplitude and GWs from individual SMBHB, and to search for anisotropies in the GWB. \citet{ltm15} computed a robust upper limit on the GWB amplitude of $A<3.0\times10^{-15}$, taking into account the presence of other spatially-correlated noise \citep{thk16}. \citet{bps15} shows that the highest sensitivity to resolvable sources is reached by EPTA between $5$ and $7\times 10^{-9}$ Hz, with a strain amplitude limit at $95\%$ between $6$ and $14\times 10^{-15}$. 
\citet{tmg15} assessed that the current EPTA dataset cannot constrain the angular distribution of the anisotropies yet, but their amplitude is $~40\%$ of the effect given by the isotropic GWB. \\ The dataset presented in \citet{dcl16} and complemented with historical data was used to carry out individual-pulsar studies of MSPs~J1024$-$0719 \citep{bjs16}, J0613$-$0200 \citep{mjs16} and J2051$-$0827 \citep{svf16}. The EPTA project LEAP \citep{bjk16, sbj17} presented a single pulse analysis of MSP~J1713+0747 \citep{lbj16}, important to assess the impact of pulse jitter on timing precision. 

\subsection{PPTA}
\citet{rhc16} presented an extension to the first PPTA data release \citep{mhb13}, which included new timing solutions for $20$ MSPs and their red noise analysis based on a new version of the Cholesky method \citep{chc11}. This study includes the first distance to a pulsar, MSP~J0437$-$4715, measured to sub-parsec precision. A multi-frequency polarization and spectral analysis of the PPTA MSPs was presented in \citet{dhm15}, finding deviations from the models commonly-applied to study pulsar spectra and Faraday rotation in some of the pulsars. \\ Additional studies of pulse jitter \citep{sod14}, extreme scattering events \citep{cks15}, differences in measured positions between VLBI and pulsar timing studies \citep{wch17}, and variations in the pulse profiles \citep{slk16} were presented between $2014$ and $2016$. \\ The now-established technique of profile-domain pulsar timing \citep{lah14} has been expanded to include the frequency-evolution of pulse profiles \citep{lkd17} and the impact on the pulse profile due to the variable scattering effects of the IISM \citep{lkd17b}. \\ \citet{srl15} used the four most stable PPTA sources and placed the most constraining upper limit on the GWB amplitude to date, $1\times10^{-15}$. \citet{mzh16} developed a new technique to search for individual GW sources, without constrains on the waveform, and the PPTA dataset was searched for individual SMBHBs and memory bursts by, respectively, \citet{zhw14} and \citet{whc15}. No evidence for GWs was found, and the two studies placed upper limits on the amplitude of the two events. In the case of resolvable SMBHB, an upper limit of $A<1.7\times10^{-14}$ was found at $10^{-8}$ Hz, while no burst events with an amplitude lower than $2\times10^{-14}$ could have been detected in the PPTA dataset studied in \citet{whc15}.

\subsection{NANOGrav}
The NANOGrav collaboration published its latest data release in $2015$ \citep{abb15}, which included the timing solutions for $37$ MSPs obtained from datasets spanning up to nine years. New methods were developed to account for variable DM and profile evolution with frequency, and 10 pulsars were found to be affected by red noise. \citet{fpe16} measured the Shapiro delay and masses for 14 MSPs in binary systems in the NANOGrav dataset, while \citet{mnf16} studied the astrometry of the 37 sources, finding the velocity dispersions to be much smaller than for the general pulsar population. \\ Detailed analyses of the effects induced by the turbulent IISM were also carried out. \citet{lmj16} analysed the scattering contribution, and concluded that the effect on the ToA errors due to variable multi-path propagation effects is negligible. \citet{jml17} studied the DM variations, finding incompatibility with a Kolmogorov spectrum \citep{ars95} in four of the pulsars, but the discrepancies can be explained by the presence of unaccounted trends in the data.\\ Noise analyses were conducted both on short \citep{lcc16} and long timescales \citep{lcc17}, and five more pulsars in addition to those identified by \citealt{abb15} were found to be affected by red noise. \\ The 9-year NANOGrav dataset was searched for GWs \citep{abb16}. No evidence of a GWB was found, and an upper limit on the GWB amplitude was set at $1.5\times10^{-15}$. The previous NANOGrav dataset \citep{dfg13} was also searched for GWs from individual sources, in the form of continuous GW emission \citep{abb14} and memory bursts \citep{abb15c}. No evidence for GWs was found, but upper limits were placed on the amplitude of continuous waves ($A<3.0\times10^{-14}$ at $10^{-8}$ Hz) and for the occurrence rate of memory bursts depending on their amplitude (e.g., memory bursts with an amplitude larger than $4\times10^{-14}$ at $6.2$ yr$^{-1}$).

\subsection{IPTA}
\begin{figure}
\begin{center}
\includegraphics[scale=0.41,angle=-90]{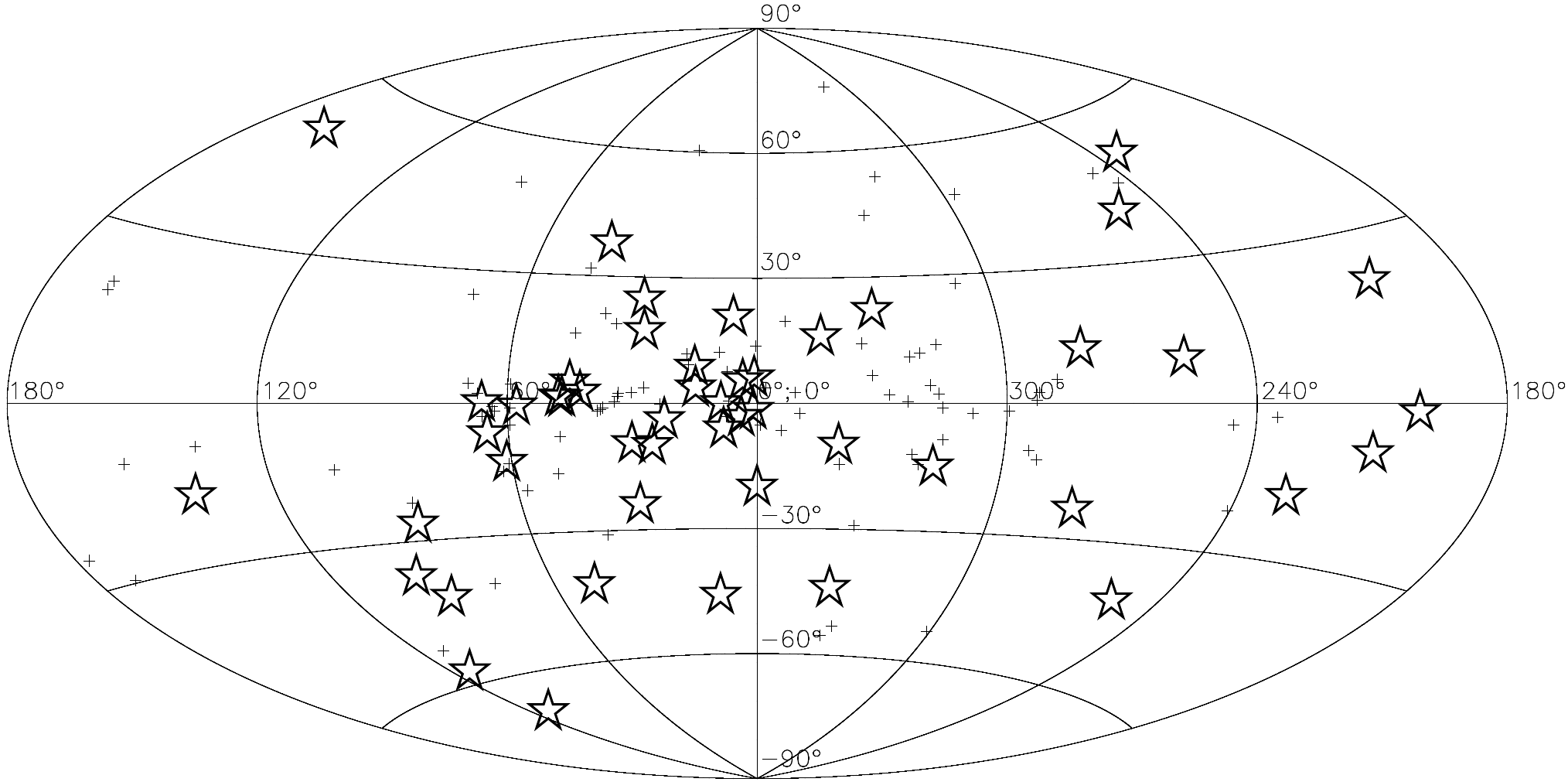}
\caption{Figure taken from \citet{vlh16}\protect\footnotemark; Aitoff projection of the IPTA MSPs. The two axes represent Galactic longitude (l) and latitude (b), while the stars represent the position of the IPTA MSPs. The crosses represent the MSPs that have been detected at radio wavelengths which are not part of a globular cluster present in the ATNF Pulsar Catalogue at the time of writing.}
\label{fig:IPTA_aitoff}
\end{center}
\end{figure}

\citet{vlh16} and \citet{lsc16} presented the first IPTA data release, based on the combination of the EPTA, PPTA, and NANOGrav datasets for $49$ MSPs (see Figure~\ref{fig:IPTA_aitoff}). The IPTA dataset consists of the ToAs time series, timing solutions, and noise models for the $49$ sources. 
The noise analysis carried out by \citet{lsc16} showed that the two main sources of red noise are variable DM, and intrinsic timing noise. However, these two sources of noise are often indistinguishable, due to a lack of multifrequency data. \\ A basic search for a GWB was carried out in \citet{vlh16}, using all of the pulsars in the array. No evidence for GWs was found, and the IPTA placed an upper limit on the GWB amplitude of $1.7\times10^{-15}$. This value, higher than the most stringent upper limit from PTA experiments ($1\times10^{-15}$, \citealt{srl15}), is more constraining than that obtained by the individual PTAs. This indicates that the IPTA as a whole is more sensitive than the individual PTAs by at least a factor of two \citep{vlh16}. \\

\footnotetext{Figure 1 of \textit{The International Pulsar Timing Array: First data release}, by Verbiest et al. 2016 (Monthly Notices of the Royal Astronomical Society, Volume 458, Issue 2, p.1267-1288) -- reused by permission of Oxford University Press. This figure is not covered by the Open-Access licence of this publication. For permissions contact \url{Journals.permissions@OUP.com}}

\section{Future prospects}\label{sec:future}

The main challenge of PTAs progresses lies in identifying and correcting for corrupting effects on the ToAs, many of which were of no importance until recently. As far as current efforts go, the bulk of the ongoing research is dedicated to study several long-period processes affecting the residuals -- effects such as inaccuracies in the Solar System ephemerides, the IISM, intrinsic pulsar-timing noise, and instrumental instabilities. In addition to those, there are continuous efforts to increase the number of highly precise MSPs in the arrays, decreasing the levels of white noise, increasing observing baselines, cadence and frequency coverage, improving analysis methods for multi-frequency data, tackling previously intractable issues visible in unprecedented high-S/N data.\\
The new generation of radio telescopes that are now coming online will greatly increase our sensitivity to low-frequency GWs. The most sensitive instrument will be the Square Kilometer Array (SKA, \citealt{bbg15}). The SKA will be built in Western Australia (low-band antennae) and in South Africa (mid- and high-band antennae), and will boost the sensitivity beyond the limits currently set by radio telescopes. The predicted probability of a GW detection after five years of observations with a SKA-based PTA (even without taking into account the current IPTA dataset, and assuming the original SKA design) is 50\% \citep{jhm15}.\\
In preparation for the SKA, several pathfinders have been deployed, such as MeerKAT and the MWA (Western Australia, \citealt{tgb13}), LWA (New Mexico, USA, \citealt{ecc09}), and the LOw Frequency ARray (Europe, \citealt{vwg13}), and they are proving to be vital to tackle some of the mentioned main challenges. For example, the low frequency facilities such as LOFAR, LWA and MWA, are fundamental instruments to monitor the turbulent IISM and its effects on pulsar timing, due to the frequency dependence of IISM effects on the propagation of radio waves. In particular, DM variations are one of the main sources of red noise in the ToA time series. IISM studies at low frequencies will be able to provide invaluable insights to improve the red noise models, and to disentangle the IISM contribution from intrinsic timing noise generated from instabilities in the pulsar spin.

\begin{acknowledgements}
The author is very grateful to Joris Verbiest, Golam Shaifullah, James McKee, Alberto Sesana, Chiara Mingarelli and Dominik Schwarz, for their availability and helpfulness, for the useful conversations and proofreading.
\end{acknowledgements}

\bibliographystyle{pasa-mnras}

\DeclareRobustCommand{\VAN}[3]{#3}

\bibliography{journals,modrefs,psrrefs,crossrefs}

\begin{thebibliography}{}
\makeatletter
\relax
\def\mn@urlcharsother{\let\do\@makeother \do\$\do\&\do\#\do\^\do\_\do\%\do\~}
\definecolor{darkblue}{rgb}{0,0,0.597656}
\def\mndoi{\begingroup\mn@urlcharsother \@ifnextchar [ {\mndoi@} {\mndoi@[]}}
\def\mndoi@[#1]#2{\def\@tempa{#1}\ifx\@tempa\@empty \href
  {http://dx.doi.org/#2} {\textcolor{darkblue}{doi:#2}}\else \href
  {http://dx.doi.org/#2} {\textcolor{darkblue}{#1}}\fi \endgroup}
\def\mn@eprint#1#2{\mn@eprint@#1:#2::\@nil}
\def\mn@eprint@arXiv#1{\href {http://arxiv.org/abs/#1} {{\tt arXiv:#1}}}
\def\mn@eprint@dblp#1{\href {http://dblp.uni-trier.de/rec/bibtex/#1.xml}
  {dblp:#1}}
\def\mn@eprint@#1:#2:#3:#4\@nil{\def\@tempa {#1}\def\@tempb {#2}\def\@tempc
  {#3}\ifx \@tempc \@empty \let \@tempc \@tempb \let \@tempb \@tempa \fi \ifx
  \@tempb \@empty \def\@tempb {arXiv}\fi \@ifundefined
  {mn@eprint@\@tempb}{\@tempb:\@tempc}{\expandafter \expandafter \csname
  mn@eprint@\@tempb\endcsname \expandafter{\@tempc}}}

\bibitem[\protect\citeauthoryear{{Aasi} et~al.,}{{Aasi} et~al.}{2013}]{aaa13}
{Aasi} J.,  et~al., 2013, \mndoi [Nature Photonics] {10.1038/nphoton.2013.177},
  7, 613

\bibitem[\protect\citeauthoryear{{Abbott} et~al.,}{{Abbott}
  et~al.}{2016a}]{aaa16}
{Abbott} B.~P.,  et~al., 2016a, \mndoi [Physical Review Letters]
  {10.1103/PhysRevLett.116.061102}, \href
  {http://adsabs.harvard.edu/abs/2016PhRvL.116f1102A} {116, 061102}

\bibitem[\protect\citeauthoryear{{Abbott} et~al.,}{{Abbott}
  et~al.}{2016b}]{aaa16b}
{Abbott} B.~P.,  et~al., 2016b, \mndoi [Physical Review Letters]
  {10.1103/PhysRevLett.116.241103}, \href
  {http://adsabs.harvard.edu/abs/2016PhRvL.116x1103A} {116, 241103}

\bibitem[\protect\citeauthoryear{{Abbott} et~al.,}{{Abbott}
  et~al.}{2017}]{aaa17b}
{Abbott} B.~P.,  et~al., 2017, \mndoi [Physical Review Letters]
  {10.1103/PhysRevLett.118.221101}, \href
  {http://adsabs.harvard.edu/abs/2017PhRvL.118v1101A} {118, 221101}

\bibitem[\protect\citeauthoryear{{Acernese} et~al.,}{{Acernese}
  et~al.}{2015}]{aaa15}
{Acernese} F.,  et~al., 2015, \mndoi [Classical and Quantum Gravity]
  {10.1088/0264-9381/32/2/024001}, \href
  {http://adsabs.harvard.edu/abs/2015CQGra..32b4001A} {32, 024001}

\bibitem[\protect\citeauthoryear{{Allen} \& {Romano}}{{Allen} \&
  {Romano}}{1999}]{ar99}
{Allen} B.,  {Romano} J.~D.,  1999, \mndoi [\prd] {10.1103/PhysRevD.59.102001},
  \href {http://adsabs.harvard.edu/abs/1999PhRvD..59j2001A} {59, 102001}

\bibitem[\protect\citeauthoryear{Alpar, Cheng, Ruderman  \& Shaham}{Alpar
  et~al.}{1982}]{acrs82}
Alpar M.~A.,  Cheng A.~F.,  Ruderman M.~A.,   Shaham J.,  1982, \mndoi [Nature]
  {10.1038/300728a0}, \href {http://adsabs.harvard.edu/abs/1982Natur.300..728A}
  {300, 728}

\bibitem[\protect\citeauthoryear{{Amaro-Seoane} et~al.,}{{Amaro-Seoane}
  et~al.}{2017}]{asa17}
{Amaro-Seoane} P.,  et~al., 2017, preprint (\mn@eprint {arXiv} {1702.00786})

\bibitem[\protect\citeauthoryear{{Araya} et~al.,}{{Araya} et~al.}{2017}]{atm17}
{Araya} A.,  et~al., 2017, \mndoi [Earth, Planets, and Space]
  {10.1186/s40623-017-0660-0}, \href
  {http://adsabs.harvard.edu/abs/2017EP%26S...69...77A} {69, 77}

\bibitem[\protect\citeauthoryear{{Armano} et~al.,}{{Armano}
  et~al.}{2017}]{aaa17}
{Armano} M.,  et~al., 2017, \mndoi [Physical Review Letters]
  {10.1103/PhysRevLett.118.171101}, \href
  {http://adsabs.harvard.edu/abs/2017PhRvL.118q1101A} {118, 171101}

\bibitem[\protect\citeauthoryear{Armstrong, Rickett  \& Spangler}{Armstrong
  et~al.}{1995}]{ars95}
Armstrong J.~W.,  Rickett B.~J.,   Spangler S.~R.,  1995, \mndoi [ApJ]
  {10.1086/175515}, \href {http://adsabs.harvard.edu/abs/1995ApJ...443..209A}
  {443, 209}

\bibitem[\protect\citeauthoryear{{Arzoumanian} et~al.,}{{Arzoumanian}
  et~al.}{2014}]{abb14}
{Arzoumanian} Z.,  et~al., 2014, \mndoi [\apj] {10.1088/0004-637X/794/2/141},
  \href {http://adsabs.harvard.edu/abs/2014ApJ...794..141A} {794, 141}

\bibitem[\protect\citeauthoryear{{Arzoumanian} et~al.,}{{Arzoumanian}
  et~al.}{2015a}]{abb15c}
{Arzoumanian} Z.,  et~al., 2015a, \mndoi [\apj] {10.1088/0004-637X/810/2/150},
  \href {http://adsabs.harvard.edu/abs/2015ApJ...810..150A} {810, 150}

\bibitem[\protect\citeauthoryear{{Arzoumanian} et~al.,}{{Arzoumanian}
  et~al.}{2015b}]{abb15}
{Arzoumanian} Z.,  et~al., 2015b, \mndoi [\apj] {10.1088/0004-637X/813/1/65},
  \href {http://adsabs.harvard.edu/abs/2015ApJ...813...65T} {813, 65}

\bibitem[\protect\citeauthoryear{{Arzoumanian} et~al.,}{{Arzoumanian}
  et~al.}{2016}]{abb16}
{Arzoumanian} Z.,  et~al., 2016, \mndoi [\apj] {10.3847/0004-637X/821/1/13},
  \href {http://adsabs.harvard.edu/abs/2016ApJ...821...13A} {821, 13}

\bibitem[\protect\citeauthoryear{{Babak} et~al.,}{{Babak} et~al.}{2016}]{bps15}
{Babak} S.,  et~al., 2016, \mndoi [\mnras] {10.1093/mnras/stv2092}, \href
  {http://adsabs.harvard.edu/abs/2016MNRAS.455.1665B} {455, 1665}

\bibitem[\protect\citeauthoryear{{Bassa} et~al.,}{{Bassa}
  et~al.}{2016a}]{bjk16}
{Bassa} C.~G.,  et~al., 2016a, \mndoi [\mnras] {10.1093/mnras/stv2755}, \href
  {http://adsabs.harvard.edu/abs/2016MNRAS.456.2196B} {456, 2196}

\bibitem[\protect\citeauthoryear{{Bassa} et~al.,}{{Bassa}
  et~al.}{2016b}]{bjs16}
{Bassa} C.~G.,  et~al., 2016b, \mndoi [\mnras] {10.1093/mnras/stw1134}, \href
  {http://adsabs.harvard.edu/abs/2016MNRAS.460.2207B} {460, 2207}

\bibitem[\protect\citeauthoryear{{Begelman}, {Blandford}  \& {Rees}}{{Begelman}
  et~al.}{1980}]{bbr80}
{Begelman} M.~C.,  {Blandford} R.~D.,   {Rees} M.~J.,  1980, \mndoi [\nat]
  {10.1038/287307a0}, \href {http://adsabs.harvard.edu/abs/1980Natur.287..307B}
  {287, 307}

\bibitem[\protect\citeauthoryear{{Benson} et~al.,}{{Benson}
  et~al.}{2014}]{baa14}
{Benson} B.~A.,  et~al., 2014, in Millimeter, Submillimeter, and Far-Infrared
  Detectors and Instrumentation for Astronomy VII. p. 91531P (\mn@eprint
  {arXiv} {1407.2973}), \mndoi{10.1117/12.2057305}

\bibitem[\protect\citeauthoryear{Bhattacharya \& {van den Heuvel}}{Bhattacharya
  \& {van den Heuvel}}{1991}]{bv91}
Bhattacharya D.,  {van den Heuvel} E. P.~J.,  1991, Phys. Rep., 203, 1

\bibitem[\protect\citeauthoryear{{Booth}, {de Blok}, {Jonas}  \&
  {Fanaroff}}{{Booth} et~al.}{2009}]{bdb09}
{Booth} R.~S.,  {de Blok} W.~J.~G.,  {Jonas} J.~L.,   {Fanaroff} B.,  2009,
  preprint, \href {http://adsabs.harvard.edu/abs/2009arXiv0910.2935B} {}
  (\mn@eprint {arXiv} {0910.2935})

\bibitem[\protect\citeauthoryear{{Braginskii} \& {Thorne}}{{Braginskii} \&
  {Thorne}}{1987}]{bt87}
{Braginskii} V.~B.,  {Thorne} K.~S.,  1987, \mndoi [nat] {10.1038/327123a0},
  \href {http://adsabs.harvard.edu/abs/1987Natur.327..123B} {327, 123}

\bibitem[\protect\citeauthoryear{{Braun}, {Bourke}, {Green}, {Keane}  \&
  {Wagg}}{{Braun} et~al.}{2015}]{bbg15}
{Braun} R.,  {Bourke} T.,  {Green} J.~A.,  {Keane} E.,   {Wagg} J.,  2015,
  Advancing Astrophysics with the Square Kilometre Array (AASKA14), p.~174

\bibitem[\protect\citeauthoryear{{Caballero} et~al.,}{{Caballero}
  et~al.}{2016}]{cll16}
{Caballero} R.~N.,  et~al., 2016, \mndoi [\mnras] {10.1093/mnras/stw179}, \href
  {http://adsabs.harvard.edu/abs/2016MNRAS.457.4421C} {457, 4421}

\bibitem[\protect\citeauthoryear{{Chamel} \& {Haensel}}{{Chamel} \&
  {Haensel}}{2008}]{ch08}
{Chamel} N.,  {Haensel} P.,  2008, \mndoi [Living Reviews in Relativity]
  {10.12942/lrr-2008-10}, \href
  {http://adsabs.harvard.edu/abs/2008LRR....11...10C} {11, 10}

\bibitem[\protect\citeauthoryear{{Champion} et~al.,}{{Champion}
  et~al.}{2010}]{chm10}
{Champion} D.~J.,  et~al., 2010, \mndoi [\apjl] {10.1088/2041-8205/720/2/L201},
  \href {http://adsabs.harvard.edu/abs/2010ApJ...720L.201C} {720, L201}

\bibitem[\protect\citeauthoryear{{Coles}, {Hobbs}, {Champion}, {Manchester}  \&
  {Verbiest}}{{Coles} et~al.}{2011}]{chc11}
{Coles} W.,  {Hobbs} G.,  {Champion} D.~J.,  {Manchester} R.~N.,   {Verbiest}
  J.~P.~W.,  2011, \mndoi [\mnras] {10.1111/j.1365-2966.2011.19505.x}, \href
  {http://adsabs.harvard.edu/abs/2011MNRAS.418..561C} {418, 561}

\bibitem[\protect\citeauthoryear{{Coles} et~al.,}{{Coles} et~al.}{2015}]{cks15}
{Coles} W.~A.,  et~al., 2015, \mndoi [\apj] {10.1088/0004-637X/808/2/113},
  \href {http://adsabs.harvard.edu/abs/2015ApJ...808..113C} {808, 113}

\bibitem[\protect\citeauthoryear{{Cordes} \& {Jenet}}{{Cordes} \&
  {Jenet}}{2012}]{cj12}
{Cordes} J.~M.,  {Jenet} F.~A.,  2012, \mndoi [\apj]
  {10.1088/0004-637X/752/1/54}, \href
  {http://adsabs.harvard.edu/abs/2012ApJ...752...54C} {752, 54}

\bibitem[\protect\citeauthoryear{{Dai} et~al.,}{{Dai} et~al.}{2015}]{dhm15}
{Dai} S.,  et~al., 2015, \mndoi [\mnras] {10.1093/mnras/stv508}, \href
  {http://adsabs.harvard.edu/abs/2015MNRAS.449.3223D} {449, 3223}

\bibitem[\protect\citeauthoryear{{Damour} \& {Vilenkin}}{{Damour} \&
  {Vilenkin}}{2005}]{dv05}
{Damour} T.,  {Vilenkin} A.,  2005, \mndoi [Phys. Rev. D]
  {10.1103/PhysRevD.71.063510}, 71, 063510

\bibitem[\protect\citeauthoryear{{Demorest} et~al.,}{{Demorest}
  et~al.}{2013}]{dfg13}
{Demorest} P.~B.,  et~al., 2013, \mndoi [\apj] {10.1088/0004-637X/762/2/94},
  \href {http://adsabs.harvard.edu/abs/2013ApJ...762...94D} {762, 94}

\bibitem[\protect\citeauthoryear{{Desvignes} et~al.,}{{Desvignes}
  et~al.}{2016}]{dcl16}
{Desvignes} G.,  et~al., 2016, \mndoi [\mnras] {10.1093/mnras/stw483}, \href
  {http://adsabs.harvard.edu/abs/2016MNRAS.458.3341D} {458, 3341}

\bibitem[\protect\citeauthoryear{Detweiler}{Detweiler}{1979}]{det79}
Detweiler S.,  1979, ApJ, 234, 1100

\bibitem[\protect\citeauthoryear{Downs}{Downs}{1981}]{dow81}
Downs G.~S.,  1981, ApJ, 249, 687

\bibitem[\protect\citeauthoryear{{DuPlain}, {Ransom}, {Demorest}, {Brandt},
  {Ford}  \& {Shelton}}{{DuPlain} et~al.}{2008}]{dpr08}
{DuPlain} R.,  {Ransom} S.,  {Demorest} P.,  {Brandt} P.,  {Ford} J.,
  {Shelton} A.~L.,  2008, in Advanced Software and Control for Astronomy II. p.
  70191D, \mndoi{10.1117/12.790003}

\bibitem[\protect\citeauthoryear{{Edwards}, {Hobbs}  \& {Manchester}}{{Edwards}
  et~al.}{2006}]{ehm06}
{Edwards} R.~T.,  {Hobbs} G.~B.,   {Manchester} R.~N.,  2006, \mndoi [MNRAS]
  {10.1111/j.1365-2966.2006.10870.x}, 372, 1549

\bibitem[\protect\citeauthoryear{Einstein}{Einstein}{1916}]{ein16}
Einstein A.,  1916, N{\"a}herungsweise Integration der Feldgleichungen der
  Gravitation

\bibitem[\protect\citeauthoryear{{Ellingson}, {Clarke}, {Cohen}, {Craig},
  {Kassim}, {Pihlstrom}, {Rickard}  \& {Taylor}}{{Ellingson}
  et~al.}{2009}]{ecc09}
{Ellingson} S.~W.,  {Clarke} T.~E.,  {Cohen} A.,  {Craig} J.,  {Kassim} N.~E.,
  {Pihlstrom} Y.,  {Rickard} L.~J.,   {Taylor} G.~B.,  2009, \mndoi [IEEE
  Proceedings] {10.1109/JPROC.2009.2015683}, \href
  {http://adsabs.harvard.edu/abs/2009IEEEP..97.1421E} {97, 1421}

\bibitem[\protect\citeauthoryear{{Favata}}{{Favata}}{2009}]{fav09}
{Favata} M.,  2009, \mndoi [\apjl] {10.1088/0004-637X/696/2/L159}, \href
  {http://adsabs.harvard.edu/abs/2009ApJ...696L.159F} {696, L159}

\bibitem[\protect\citeauthoryear{{Flanagan} \& {Hughes}}{{Flanagan} \&
  {Hughes}}{1998}]{fh98}
{Flanagan} {\'E}.~{\'E}.,  {Hughes} S.~A.,  1998, \mndoi [\prd]
  {10.1103/PhysRevD.57.4535}, \href
  {http://adsabs.harvard.edu/abs/1998PhRvD..57.4535F} {57, 4535}

\bibitem[\protect\citeauthoryear{{Fonseca} et~al.,}{{Fonseca}
  et~al.}{2016}]{fpe16}
{Fonseca} E.,  et~al., 2016, \mndoi [\apj] {10.3847/0004-637X/832/2/167}, \href
  {http://adsabs.harvard.edu/abs/2016ApJ...832..167F} {832, 167}

\bibitem[\protect\citeauthoryear{Foster \& Backer}{Foster \&
  Backer}{1990}]{fb90}
Foster R.~S.,  Backer D.~C.,  1990, \mndoi [ApJ] {10.1086/169195}, \href
  {http://adsabs.harvard.edu/abs/1990ApJ...361..300F} {361, 300}

\bibitem[\protect\citeauthoryear{{Geyer} et~al.,}{{Geyer} et~al.}{2017}]{gk17}
{Geyer} M.,  et~al., 2017, \mndoi [\mnras] {10.1093/mnras/stx1151}, \href
  {http://adsabs.harvard.edu/abs/2017MNRAS.470.2659G} {470, 2659}

\bibitem[\protect\citeauthoryear{{Grishchuk}}{{Grishchuk}}{1974}]{gri74}
{Grishchuk} L.~P.,  1974, Zhurnal Eksperimentalnoi i Teoreticheskoi Fiziki,
  \href {http://adsabs.harvard.edu/abs/1974ZhETF..67..825G} {67, 825}

\bibitem[\protect\citeauthoryear{{Grishchuk}}{{Grishchuk}}{2005}]{gri05}
{Grishchuk} L.~P.,  2005, Phys. Uspekhi, 48, 1235

\bibitem[\protect\citeauthoryear{{\VAN{Haarlem}{van}{van}}~Haarlem
  et~al.,}{{\VAN{Haarlem}{van}{van}}~Haarlem et~al.}{2013}]{vwg13}
{\VAN{Haarlem}{van}{van}}~Haarlem M.~P.,  et~al., 2013, \mndoi [\aap]
  {10.1051/0004-6361/201220873}, \href
  {http://adsabs.harvard.edu/abs/2013A%26A...556A...2V} {556, A2}

\bibitem[\protect\citeauthoryear{{\VAN{Haasteren}{van}{van}}~Haasteren \&
  {Levin}}{{\VAN{Haasteren}{van}{van}}~Haasteren \& {Levin}}{2010}]{vhl10}
{\VAN{Haasteren}{van}{van}}~Haasteren R.,  {Levin} Y.,  2010, \mndoi [\mnras]
  {10.1111/j.1365-2966.2009.15885.x}, \href
  {http://adsabs.harvard.edu/abs/2010MNRAS.401.2372V} {401, 2372}

\bibitem[\protect\citeauthoryear{Hellings \& Downs}{Hellings \&
  Downs}{1983}]{hd83}
Hellings R.~W.,  Downs G.~S.,  1983, \apjl, 265, L39

\bibitem[\protect\citeauthoryear{Hobbs, Lyne, Kramer, Martin  \& Jordan}{Hobbs
  et~al.}{2004}]{hlk+04}
Hobbs G.,  Lyne A.~G.,  Kramer M.,  Martin C.~E.,   Jordan C.,  2004, MNRAS,
  353, 1311

\bibitem[\protect\citeauthoryear{{Hobbs}, {Edwards}  \& {Manchester}}{{Hobbs}
  et~al.}{2006}]{hem06}
{Hobbs} G.~B.,  {Edwards} R.~T.,   {Manchester} R.~N.,  2006, \mndoi [MNRAS]
  {10.1111/j.1365-2966.2006.10302.x}, 369, 655

\bibitem[\protect\citeauthoryear{{Hobbs} et~al.,}{{Hobbs} et~al.}{2012}]{hcm12}
{Hobbs} G.,  et~al., 2012, \mndoi [\mnras] {10.1111/j.1365-2966.2012.21946.x},
  \href {http://adsabs.harvard.edu/abs/2012MNRAS.427.2780H} {427, 2780}

\bibitem[\protect\citeauthoryear{{Hughes}}{{Hughes}}{2003}]{hug03}
{Hughes} S.~A.,  2003, \mndoi [Annals of Physics]
  {10.1016/S0003-4916(02)00025-8}, \href
  {http://adsabs.harvard.edu/abs/2003AnPhy.303..142H} {303, 142}

\bibitem[\protect\citeauthoryear{Hulse \& Taylor}{Hulse \& Taylor}{1974}]{ht74}
Hulse R.~A.,  Taylor J.~H.,  1974, \mndoi [\apjl] {10.1086/181548}, \href
  {http://adsabs.harvard.edu/abs/1974ApJ...191L..59H} {191, L59}

\bibitem[\protect\citeauthoryear{{Janssen} et~al.,}{{Janssen}
  et~al.}{2015}]{jhm15}
{Janssen} G.,  et~al., 2015, Advancing Astrophysics with the Square Kilometre
  Array (AASKA14), \href {http://adsabs.harvard.edu/abs/2015aska.confE..37J}
  {p.~37}

\bibitem[\protect\citeauthoryear{{Jones} et~al.,}{{Jones} et~al.}{2017}]{jml17}
{Jones} M.~L.,  et~al., 2017, \mndoi [\apj] {10.3847/1538-4357/aa73df}, \href
  {http://adsabs.harvard.edu/abs/2017ApJ...841..125J} {841, 125}

\bibitem[\protect\citeauthoryear{{Karuppusamy}, {Stappers}  \& {van
  Straten}}{{Karuppusamy} et~al.}{2008}]{ksv08}
{Karuppusamy} R.,  {Stappers} B.,   {van Straten} W.,  2008, \mndoi [\pasp]
  {10.1086/528699}, \href {http://adsabs.harvard.edu/abs/2008PASP..120..191K}
  {120, 191}

\bibitem[\protect\citeauthoryear{{Keating}, {Ade}, {Bock}, {Hivon},
  {Holzapfel}, {Lange}, {Nguyen}  \& {Yoon}}{{Keating} et~al.}{2003}]{kab03}
{Keating} B.~G.,  {Ade} P.~A.~R.,  {Bock} J.~J.,  {Hivon} E.,  {Holzapfel}
  W.~L.,  {Lange} A.~E.,  {Nguyen} H.,   {Yoon} K.~W.,  2003, in {Fineschi} S.,
   ed.,  \procspie Vol. 4843, Polarimetry in Astronomy. pp 284--295,
  \mndoi{10.1117/12.459274}

\bibitem[\protect\citeauthoryear{{Kermish} et~al.,}{{Kermish}
  et~al.}{2012}]{kaa12}
{Kermish} Z.~D.,  et~al., 2012, in Millimeter, Submillimeter, and Far-Infrared
  Detectors and Instrumentation for Astronomy VI. p. 84521C (\mn@eprint {arXiv}
  {1210.7768}), \mndoi{10.1117/12.926354}

\bibitem[\protect\citeauthoryear{{Kibble}}{{Kibble}}{1976}]{kib76}
{Kibble} T.~W.~B.,  1976, \mndoi [Journal of Physics A Mathematical General]
  {10.1088/0305-4470/9/8/029}, \href
  {http://adsabs.harvard.edu/abs/1976JPhA....9.1387K} {9, 1387}

\bibitem[\protect\citeauthoryear{{Komatsu} et~al.,}{{Komatsu}
  et~al.}{2011}]{ksd11}
{Komatsu} E.,  et~al., 2011, \mndoi [\apjs] {10.1088/0067-0049/192/2/18}, \href
  {http://adsabs.harvard.edu/abs/2011ApJS..192...18K} {192, 18}

\bibitem[\protect\citeauthoryear{{Kormendy} \& {Ho}}{{Kormendy} \&
  {Ho}}{2013}]{kh13}
{Kormendy} J.,  {Ho} L.~C.,  2013, \mndoi [\araa]
  {10.1146/annurev-astro-082708-101811}, \href
  {http://adsabs.harvard.edu/abs/2013ARA%26A..51..511K} {51, 511}

\bibitem[\protect\citeauthoryear{{Kormendy} \& {Richstone}}{{Kormendy} \&
  {Richstone}}{1995}]{kr95}
{Kormendy} J.,  {Richstone} D.,  1995, \mndoi [\araa]
  {10.1146/annurev.aa.33.090195.003053}, \href
  {http://adsabs.harvard.edu/abs/1995ARA%26A..33..581K} {33, 581}

\bibitem[\protect\citeauthoryear{{Kovac}, {Leitch}, {Pryke}, {Carlstrom},
  {Halverson}  \& {Holzapfel}}{{Kovac} et~al.}{2002}]{kl02}
{Kovac} J.~M.,  {Leitch} E.~M.,  {Pryke} C.,  {Carlstrom} J.~E.,  {Halverson}
  N.~W.,   {Holzapfel} W.~L.,  2002, \mndoi [\nat] {10.1038/nature01269}, \href
  {http://adsabs.harvard.edu/abs/2002Natur.420..772K} {420, 772}

\bibitem[\protect\citeauthoryear{Kramer, Xilouris, Lorimer, Doroshenko,
  Jessner, Wielebinski, Wolszczan  \& Camilo}{Kramer et~al.}{1998}]{kxc+98}
Kramer M.,  Xilouris K.~M.,  Lorimer D.,  Doroshenko O.,  Jessner A.,
  Wielebinski R.,  Wolszczan A.,   Camilo F.,  1998, ApJ, 501, 270

\bibitem[\protect\citeauthoryear{{Kramer} et~al.,}{{Kramer}
  et~al.}{2006}]{ksm+06}
{Kramer} M.,  et~al., 2006, \mndoi [Science] {10.1126/science.1132305}, 314, 97

\bibitem[\protect\citeauthoryear{{Lam} et~al.,}{{Lam} et~al.}{2016}]{lcc16}
{Lam} M.~T.,  et~al., 2016, \mndoi [\apj] {10.3847/0004-637X/819/2/155}, \href
  {http://adsabs.harvard.edu/abs/2016ApJ...819..155L} {819, 155}

\bibitem[\protect\citeauthoryear{{Lam} et~al.,}{{Lam} et~al.}{2017}]{lcc17}
{Lam} M.~T.,  et~al., 2017, \mndoi [\apj] {10.3847/1538-4357/834/1/35}, \href
  {http://adsabs.harvard.edu/abs/2017ApJ...834...35L} {834, 35}

\bibitem[\protect\citeauthoryear{{Lasky} et~al.,}{{Lasky} et~al.}{2016}]{lms16}
{Lasky} P.~D.,  et~al., 2016, Physical Review X, 6, 011035

\bibitem[\protect\citeauthoryear{Lattimer \& Prakash}{Lattimer \&
  Prakash}{2004}]{lp04}
Lattimer J.~H.,  Prakash M.,  2004, \mndoi [Science] {10.1126/science.1090720},
  \href {http://adsabs.harvard.edu/abs/2004Sci...304..536L} {304, 536}

\bibitem[\protect\citeauthoryear{{Lazarus}, {Karuppusamy}, {Graikou},
  {Caballero}, {Champion}, {Lee}, {Verbiest}  \& {Kramer}}{{Lazarus}
  et~al.}{2016}]{lkg16}
{Lazarus} P.,  {Karuppusamy} R.,  {Graikou} E.,  {Caballero} R.~N.,  {Champion}
  D.~J.,  {Lee} K.~J.,  {Verbiest} J.~P.~W.,   {Kramer} M.,  2016, \mndoi
  [\mnras] {10.1093/mnras/stw189}, \href
  {http://adsabs.harvard.edu/abs/2016MNRAS.458..868L} {458, 868}

\bibitem[\protect\citeauthoryear{{Lee}}{{Lee}}{2016}]{lee16}
{Lee} K.~J.,  2016, in {Qain} L.,  {Li} D.,  eds,  Astronomical Society of the
  Pacific Conference Series Vol. 502, Frontiers in Radio Astronomy and FAST
  Early Sciences Symposium 2015. p.~19

\bibitem[\protect\citeauthoryear{{Lentati}, {Alexander}, {Hobson}, {Feroz},
  {van Haasteren}, {Lee}  \& {Shannon}}{{Lentati} et~al.}{2014}]{lah14}
{Lentati} L.,  {Alexander} P.,  {Hobson} M.~P.,  {Feroz} F.,  {van Haasteren}
  R.,  {Lee} K.~J.,   {Shannon} R.~M.,  2014, \mndoi [\mnras]
  {10.1093/mnras/stt2122}, \href
  {http://adsabs.harvard.edu/abs/2014MNRAS.437.3004L} {437, 3004}

\bibitem[\protect\citeauthoryear{{Lentati} et~al.,}{{Lentati}
  et~al.}{2015}]{ltm15}
{Lentati} L.,  et~al., 2015, \mndoi [\mnras] {10.1093/mnras/stv1538}, \href
  {http://adsabs.harvard.edu/abs/2015MNRAS.453.2576L} {453, 2576}

\bibitem[\protect\citeauthoryear{{Lentati} et~al.,}{{Lentati}
  et~al.}{2016}]{lsc16}
{Lentati} L.,  et~al., 2016, \mndoi [\mnras] {10.1093/mnras/stw395}, \href
  {http://adsabs.harvard.edu/abs/2016MNRAS.458.2161L} {458, 2161}

\bibitem[\protect\citeauthoryear{{Lentati} et~al.,}{{Lentati}
  et~al.}{2017a}]{lkd17}
{Lentati} L.,  et~al., 2017a, \mndoi [\mnras] {10.1093/mnras/stw3359}, \href
  {http://adsabs.harvard.edu/abs/2017MNRAS.466.3706L} {466, 3706}

\bibitem[\protect\citeauthoryear{{Lentati}, {Kerr}, {Dai}, {Shannon}, {Hobbs}
  \& {Os{\l}owski}}{{Lentati} et~al.}{2017b}]{lk17}
{Lentati} L.,  {Kerr} M.,  {Dai} S.,  {Shannon} R.~M.,  {Hobbs} G.,
  {Os{\l}owski} S.,  2017b, \mndoi [\mnras] {10.1093/mnras/stx580}, \href
  {http://adsabs.harvard.edu/abs/2017MNRAS.468.1474L} {468, 1474}

\bibitem[\protect\citeauthoryear{{Lentati}, {Kerr}, {Dai}, {Shannon}, {Hobbs}
  \& {Os{\l}owski}}{{Lentati} et~al.}{2017c}]{lkd17b}
{Lentati} L.,  {Kerr} M.,  {Dai} S.,  {Shannon} R.~M.,  {Hobbs} G.,
  {Os{\l}owski} S.,  2017c, \mndoi [\mnras] {10.1093/mnras/stx580}, \href
  {http://adsabs.harvard.edu/abs/2017MNRAS.468.1474L} {468, 1474}

\bibitem[\protect\citeauthoryear{{Levin} et~al.,}{{Levin} et~al.}{2016}]{lmj16}
{Levin} L.,  et~al., 2016, \mndoi [\apj] {10.3847/0004-637X/818/2/166}, \href
  {http://adsabs.harvard.edu/abs/2016ApJ...818..166L} {818, 166}

\bibitem[\protect\citeauthoryear{{Liu} et~al.,}{{Liu} et~al.}{2014}]{ldc14}
{Liu} K.,  et~al., 2014, \mndoi [\mnras] {10.1093/mnras/stu1420}, \href
  {http://adsabs.harvard.edu/abs/2014MNRAS.443.3752L} {443, 3752}

\bibitem[\protect\citeauthoryear{{Liu} et~al.,}{{Liu} et~al.}{2016a}]{lb16}
{Liu} K.,  et~al., 2016a, \mndoi [\mnras] {10.1093/mnras/stw2223}, \href
  {http://adsabs.harvard.edu/abs/2016MNRAS.463.3239L} {463, 3239}

\bibitem[\protect\citeauthoryear{{Liu} et~al.,}{{Liu} et~al.}{2016b}]{lbj16}
{Liu} K.,  et~al., 2016b, \mndoi [\mnras] {10.1093/mnras/stw2223}, \href
  {http://adsabs.harvard.edu/abs/2016MNRAS.463.3239L} {463, 3239}

\bibitem[\protect\citeauthoryear{{Lommen}}{{Lommen}}{2015}]{lom15}
{Lommen} A.~N.,  2015, \mndoi [Reports on Progress in Physics]
  {10.1088/0034-4885/78/12/124901}, \href
  {http://adsabs.harvard.edu/abs/2015RPPh...78l4901L} {78, 124901}

\bibitem[\protect\citeauthoryear{Lorimer \& Kramer}{Lorimer \&
  Kramer}{2005}]{lk05}
Lorimer D.~R.,  Kramer M.,  2005, Handbook of Pulsar Astronomy.
Cambridge University Press

\bibitem[\protect\citeauthoryear{Lorimer, Yates, Lyne  \& Gould}{Lorimer
  et~al.}{1995}]{lylg95}
Lorimer D.~R.,  Yates J.~A.,  Lyne A.~G.,   Gould D.~M.,  1995, MNRAS, 273, 411

\bibitem[\protect\citeauthoryear{{Lyne}, {Hobbs}, {Kramer}, {Stairs}  \&
  {Stappers}}{{Lyne} et~al.}{2010}]{lhk10}
{Lyne} A.,  {Hobbs} G.,  {Kramer} M.,  {Stairs} I.,   {Stappers} B.,  2010,
  \mndoi [Science] {10.1126/science.1186683}, \href
  {http://adsabs.harvard.edu/abs/2010Sci...329..408L} {329, 408}

\bibitem[\protect\citeauthoryear{{Madison}, {Cordes}  \&
  {Chatterjee}}{{Madison} et~al.}{2014}]{mcc14}
{Madison} D.~R.,  {Cordes} J.~M.,   {Chatterjee} S.,  2014, \mndoi [\apj]
  {10.1088/0004-637X/788/2/141}, \href
  {http://adsabs.harvard.edu/abs/2014ApJ...788..141M} {788, 141}

\bibitem[\protect\citeauthoryear{{Madison} et~al.,}{{Madison}
  et~al.}{2016}]{mzh16}
{Madison} D.~R.,  et~al., 2016, \mndoi [\mnras] {10.1093/mnras/stv2534}, \href
  {http://adsabs.harvard.edu/abs/2016MNRAS.455.3662M} {455, 3662}

\bibitem[\protect\citeauthoryear{Maggiore}{Maggiore}{2007}]{mag07}
Maggiore M.,  2007, Gravitational waves. Volume 1: Theory and experiments.
Oxford University Press

\bibitem[\protect\citeauthoryear{{Magorrian} et~al.,}{{Magorrian}
  et~al.}{1998}]{mtr98}
{Magorrian} J.,  et~al., 1998, \mndoi [\aj] {10.1086/300353}, \href
  {http://adsabs.harvard.edu/abs/1998AJ....115.2285M} {115, 2285}

\bibitem[\protect\citeauthoryear{{Manchester} et~al.,}{{Manchester}
  et~al.}{2013}]{mhb13}
{Manchester} R.~N.,  et~al., 2013, \mndoi [\pasa] {10.1017/pasa.2012.017},
  \href {http://adsabs.harvard.edu/abs/2013PASA...30...17M} {30, e017}

\bibitem[\protect\citeauthoryear{{Matthews} et~al.,}{{Matthews}
  et~al.}{2016}]{mnf16}
{Matthews} A.~M.,  et~al., 2016, \mndoi [\apj] {10.3847/0004-637X/818/1/92},
  \href {http://adsabs.harvard.edu/abs/2016ApJ...818...92M} {818, 92}

\bibitem[\protect\citeauthoryear{{McKee} et~al.,}{{McKee} et~al.}{2016}]{mjs16}
{McKee} J.~W.,  et~al., 2016, \mndoi [\mnras] {10.1093/mnras/stw1442}, \href
  {http://adsabs.harvard.edu/abs/2016MNRAS.461.2809M} {461, 2809}

\bibitem[\protect\citeauthoryear{{Mingarelli} \& {Sidery}}{{Mingarelli} \&
  {Sidery}}{2014}]{ms14}
{Mingarelli} C.~M.~F.,  {Sidery} T.,  2014, \mndoi [\prd]
  {10.1103/PhysRevD.90.062011}, \href
  {http://adsabs.harvard.edu/abs/2014PhRvD..90f2011M} {90, 062011}

\bibitem[\protect\citeauthoryear{{Peng}, {Nan}, {Su}, {Qiu}, {Zhu}  \&
  {Zhu}}{{Peng} et~al.}{2001}]{pns01}
{Peng} B.,  {Nan} R.,  {Su} Y.,  {Qiu} Y.,  {Zhu} L.,   {Zhu} W.,  2001, \mndoi
  [\apss] {10.1023/A:1013127316851}, \href
  {http://adsabs.harvard.edu/abs/2001Ap%26SS.278..219P} {278, 219}

\bibitem[\protect\citeauthoryear{{Phinney}}{{Phinney}}{2001}]{phi01}
{Phinney} E.~S.,  2001, ArXiv Astrophysics e-prints, \href
  {http://adsabs.harvard.edu/abs/2001astro.ph..8028P} {}

\bibitem[\protect\citeauthoryear{{Planck Collaboration} et~al.,}{{Planck
  Collaboration} et~al.}{2014}]{pla14}
{Planck Collaboration} et~al., 2014, \mndoi [\aap]
  {10.1051/0004-6361/201321529}, \href
  {http://adsabs.harvard.edu/abs/2014A%26A...571A...1P} {571, A1}

\bibitem[\protect\citeauthoryear{{Polnarev}}{{Polnarev}}{1985}]{pol85}
{Polnarev} A.~G.,  1985, \azh, \href
  {http://adsabs.harvard.edu/abs/1985AZh....62.1041P} {62, 1041}

\bibitem[\protect\citeauthoryear{{Rajagopal} \& {Romani}}{{Rajagopal} \&
  {Romani}}{1995}]{rr95a}
{Rajagopal} M.,  {Romani} R.~W.,  1995, \mndoi [ApJ] {10.1086/175813}, \href
  {http://adsabs.harvard.edu/abs/1995ApJ...446..543R} {446, 543}

\bibitem[\protect\citeauthoryear{{Ravi}, {Wyithe}, {Shannon}  \&
  {Hobbs}}{{Ravi} et~al.}{2015}]{rwsh15}
{Ravi} V.,  {Wyithe} J.~S.~B.,  {Shannon} R.~M.,   {Hobbs} G.,  2015, \mndoi
  [\mnras] {10.1093/mnras/stu2659}, \href
  {http://adsabs.harvard.edu/abs/2015MNRAS.447.2772R} {447, 2772}

\bibitem[\protect\citeauthoryear{{Reardon} et~al.,}{{Reardon}
  et~al.}{2016}]{rhc16}
{Reardon} D.~J.,  et~al., 2016, \mndoi [\mnras] {10.1093/mnras/stv2395}, \href
  {http://adsabs.harvard.edu/abs/2016MNRAS.455.1751R} {455, 1751}

\bibitem[\protect\citeauthoryear{{Romani}}{{Romani}}{1989}]{rom89}
{Romani} R.~W.,  1989, in {{\"O}gelman} H.,  {van den Heuvel} E.~P.~J.,  eds,
  Timing Neutron Stars. pp 113--117

\bibitem[\protect\citeauthoryear{{Rosado}, {Sesana}  \& {Gair}}{{Rosado}
  et~al.}{2015}]{rsg15}
{Rosado} P.~A.,  {Sesana} A.,   {Gair} J.,  2015, \mndoi [\mnras]
  {10.1093/mnras/stv1098}, \href
  {http://adsabs.harvard.edu/abs/2015MNRAS.451.2417R} {451, 2417}

\bibitem[\protect\citeauthoryear{{Sazhin}}{{Sazhin}}{1978}]{saz78}
{Sazhin} M.~V.,  1978, \sovast, \href
  {http://adsabs.harvard.edu/abs/1978SvA....22...36S} {22, 36}

\bibitem[\protect\citeauthoryear{{Sesana}}{{Sesana}}{2013a}]{ses13a}
{Sesana} A.,  2013a, \mndoi [Classical and Quantum Gravity]
  {10.1088/0264-9381/30/24/244009}, \href
  {http://adsabs.harvard.edu/abs/2013CQGra..30x4009S} {30, 244009}

\bibitem[\protect\citeauthoryear{{Sesana}}{{Sesana}}{2013b}]{ses13b}
{Sesana} A.,  2013b, \mndoi [\mnras] {10.1093/mnrasl/slt034}, \href
  {http://adsabs.harvard.edu/abs/2013MNRAS.433L...1S} {433, L1}

\bibitem[\protect\citeauthoryear{{Sesana} \& {Vecchio}}{{Sesana} \&
  {Vecchio}}{2010}]{sv10}
{Sesana} A.,  {Vecchio} A.,  2010, \mndoi [\prd] {10.1103/PhysRevD.81.104008},
  \href {http://adsabs.harvard.edu/abs/2010PhRvD..81j4008S} {81, 104008}

\bibitem[\protect\citeauthoryear{{Sesana}, {Haardt}, {Madau}  \&
  {Volonteri}}{{Sesana} et~al.}{2004}]{shmv04}
{Sesana} A.,  {Haardt} F.,  {Madau} P.,   {Volonteri} M.,  2004, \mndoi [ApJ]
  {10.1086/422185}, 611, 623

\bibitem[\protect\citeauthoryear{{Sesana}, {Vecchio}  \& {Colacino}}{{Sesana}
  et~al.}{2008}]{svc08}
{Sesana} A.,  {Vecchio} A.,   {Colacino} C.~N.,  2008, \mndoi [MNRAS]
  {10.1111/j.1365-2966.2008.13682.x}, \href
  {http://adsabs.harvard.edu/abs/2008MNRAS.390..192S} {390, 192}

\bibitem[\protect\citeauthoryear{{Sesana}, {Shankar}, {Bernardi}  \&
  {Sheth}}{{Sesana} et~al.}{2016}]{ssb16}
{Sesana} A.,  {Shankar} F.,  {Bernardi} M.,   {Sheth} R.~K.,  2016, \mndoi
  [\mnras] {10.1093/mnrasl/slw139}, \href
  {http://adsabs.harvard.edu/abs/2016MNRAS.463L...6S} {463, L6}

\bibitem[\protect\citeauthoryear{{Shaifullah} et~al.,}{{Shaifullah}
  et~al.}{2016}]{svf16}
{Shaifullah} G.,  et~al., 2016, \mndoi [\mnras] {10.1093/mnras/stw1737}, \href
  {http://adsabs.harvard.edu/abs/2016MNRAS.462.1029S} {462, 1029}

\bibitem[\protect\citeauthoryear{{Shankar} et~al.,}{{Shankar}
  et~al.}{2016}]{sbs16}
{Shankar} F.,  et~al., 2016, \mndoi [\mnras] {10.1093/mnras/stw678}, \href
  {http://adsabs.harvard.edu/abs/2016MNRAS.460.3119S} {460, 3119}

\bibitem[\protect\citeauthoryear{{Shannon} \& {Cordes}}{{Shannon} \&
  {Cordes}}{2010}]{sc10}
{Shannon} R.~M.,  {Cordes} J.~M.,  2010, \mndoi [ApJ]
  {10.1088/0004-637X/725/2/1607}, \href
  {http://adsabs.harvard.edu/abs/2010ApJ...725.1607S} {725, 1607}

\bibitem[\protect\citeauthoryear{{Shannon} et~al.,}{{Shannon}
  et~al.}{2014}]{sod14}
{Shannon} R.~M.,  et~al., 2014, \mndoi [\mnras] {10.1093/mnras/stu1213}, \href
  {http://adsabs.harvard.edu/abs/2014MNRAS.443.1463S} {443, 1463}

\bibitem[\protect\citeauthoryear{{Shannon} et~al.,}{{Shannon}
  et~al.}{2015}]{srl15}
{Shannon} R.~M.,  et~al., 2015, \mndoi [Science] {10.1126/science.aab1910},
  \href {http://adsabs.harvard.edu/abs/2015Sci...349.1522S} {349, 1522}

\bibitem[\protect\citeauthoryear{{Shannon} et~al.,}{{Shannon}
  et~al.}{2016}]{slk16}
{Shannon} R.~M.,  et~al., 2016, \mndoi [\apjl] {10.3847/2041-8205/828/1/L1},
  \href {http://adsabs.harvard.edu/abs/2016ApJ...828L...1S} {828, L1}

\bibitem[\protect\citeauthoryear{{Shao}, {Caballero}, {Kramer}, {Wex},
  {Champion}  \& {Jessner}}{{Shao} et~al.}{2013}]{sck13}
{Shao} L.,  {Caballero} R.~N.,  {Kramer} M.,  {Wex} N.,  {Champion} D.~J.,
  {Jessner} A.,  2013, \mndoi [Classical and Quantum Gravity]
  {10.1088/0264-9381/30/16/165019}, \href
  {http://adsabs.harvard.edu/abs/2013CQGra..30p5019S} {30, 165019}

\bibitem[\protect\citeauthoryear{{Smits} et~al.,}{{Smits} et~al.}{2017}]{sbj17}
{Smits} R.,  et~al., 2017, Astronomy and Computing, 19, 66

\bibitem[\protect\citeauthoryear{Stairs}{Stairs}{2003}]{sta03}
Stairs I.~H.,  2003, Living Reviews in Relativity, 5

\bibitem[\protect\citeauthoryear{{Starobinski{\v i}}}{{Starobinski{\v
  i}}}{1979}]{sta79}
{Starobinski{\v i}} A.~A.,  1979, Soviet Journal of Experimental and
  Theoretical Physics Letters, \href
  {http://adsabs.harvard.edu/abs/1979JETPL..30..682S} {30, 682}

\bibitem[\protect\citeauthoryear{Taylor}{Taylor}{1992}]{tay92}
Taylor J.~H.,  1992, Philos. Trans. Roy. Soc. London A, 341, 117

\bibitem[\protect\citeauthoryear{{Taylor} et~al.,}{{Taylor}
  et~al.}{2015}]{tmg15}
{Taylor} S.~R.,  et~al., 2015, \mndoi [Physical Review Letters]
  {10.1103/PhysRevLett.115.041101}, \href
  {http://adsabs.harvard.edu/abs/2015PhRvL.115d1101T} {115, 041101}

\bibitem[\protect\citeauthoryear{{Taylor}, {Lentati}, {Babak}, {Brem}, {Gair},
  {Sesana}  \& {Vecchio}}{{Taylor} et~al.}{2017}]{tlb17}
{Taylor} S.~R.,  {Lentati} L.,  {Babak} S.,  {Brem} P.,  {Gair} J.~R.,
  {Sesana} A.,   {Vecchio} A.,  2017, \prd, 95, 042002

\bibitem[\protect\citeauthoryear{Thorne}{Thorne}{1987}]{tho87}
Thorne K.~S.,  1987, in Hawking S.,  Israel W.,  eds, 300 Years of Gravitation.
  Cambridge University Press, Cambridge, pp 330--458

\bibitem[\protect\citeauthoryear{{Tiburzi} et~al.,}{{Tiburzi}
  et~al.}{2016}]{thk16}
{Tiburzi} C.,  et~al., 2016, \mndoi [\mnras] {10.1093/mnras/stv2143}, \href
  {http://adsabs.harvard.edu/abs/2016MNRAS.455.4339T} {455, 4339}

\bibitem[\protect\citeauthoryear{{Tingay} et~al.,}{{Tingay}
  et~al.}{2013}]{tgb13}
{Tingay} S.~J.,  et~al., 2013, \mndoi [\pasa] {10.1017/pasa.2012.007}, \href
  {http://adsabs.harvard.edu/abs/2013PASA...30....7T} {30, e007}

\bibitem[\protect\citeauthoryear{{Tucci}, {Mart{\'{\i}}nez-Gonz{\'a}lez},
  {Vielva}  \& {Delabrouille}}{{Tucci} et~al.}{2005}]{tmv05}
{Tucci} M.,  {Mart{\'{\i}}nez-Gonz{\'a}lez} E.,  {Vielva} P.,   {Delabrouille}
  J.,  2005, \mndoi [\mnras] {10.1111/j.1365-2966.2005.09123.x}, \href
  {http://adsabs.harvard.edu/abs/2005MNRAS.360..935T} {360, 935}

\bibitem[\protect\citeauthoryear{{Vecchio}}{{Vecchio}}{2004}]{vec04}
{Vecchio} A.,  2004, \mndoi [\prd] {10.1103/PhysRevD.70.042001}, \href
  {http://adsabs.harvard.edu/abs/2004PhRvD..70d2001V} {70, 042001}

\bibitem[\protect\citeauthoryear{{Verbiest} et~al.,}{{Verbiest}
  et~al.}{2009}]{vbc09}
{Verbiest} J.~P.~W.,  et~al., 2009, \mndoi [\mnras]
  {10.1111/j.1365-2966.2009.15508.x}, \href
  {http://adsabs.harvard.edu/abs/2009MNRAS.400..951V} {400, 951}

\bibitem[\protect\citeauthoryear{{Verbiest} et~al.,}{{Verbiest}
  et~al.}{2016}]{vlh16}
{Verbiest} J.~P.~W.,  et~al., 2016, \mndoi [\mnras] {10.1093/mnras/stw347},
  \href {http://adsabs.harvard.edu/abs/2016MNRAS.458.1267V} {458, 1267}

\bibitem[\protect\citeauthoryear{{Volonteri}, {Haardt}  \& {Madau}}{{Volonteri}
  et~al.}{2003}]{vhm03}
{Volonteri} M.,  {Haardt} F.,   {Madau} P.,  2003, \mndoi [\apj]
  {10.1086/344675}, \href {http://adsabs.harvard.edu/abs/2003ApJ...582..559V}
  {582, 559}

\bibitem[\protect\citeauthoryear{{Wang} et~al.,}{{Wang} et~al.}{2015}]{whc15}
{Wang} J.~B.,  et~al., 2015, \mndoi [\mnras] {10.1093/mnras/stu2137}, \href
  {http://adsabs.harvard.edu/abs/2015MNRAS.446.1657W} {446, 1657}

\bibitem[\protect\citeauthoryear{{Wang} et~al.,}{{Wang} et~al.}{2017}]{wch17}
{Wang} J.~B.,  et~al., 2017, \mndoi [\mnras] {10.1093/mnras/stx837}, \href
  {http://adsabs.harvard.edu/abs/2017MNRAS.469..425W} {469, 425}

\bibitem[\protect\citeauthoryear{{Weisberg} \& {Huang}}{{Weisberg} \&
  {Huang}}{2016}]{wh16}
{Weisberg} J.~M.,  {Huang} Y.,  2016, \mndoi [\apj]
  {10.3847/0004-637X/829/1/55}, \href
  {http://adsabs.harvard.edu/abs/2016ApJ...829...55W} {829, 55}

\bibitem[\protect\citeauthoryear{Weisberg \& Taylor}{Weisberg \&
  Taylor}{1981}]{wt81}
Weisberg J.,  Taylor J.,  1981, Gen. Relativ. Gravit., 13, 1

\bibitem[\protect\citeauthoryear{{Weisberg}, {Nice}  \& {Taylor}}{{Weisberg}
  et~al.}{2010}]{wnt10}
{Weisberg} J.~M.,  {Nice} D.~J.,   {Taylor} J.~H.,  2010, \mndoi [ApJ]
  {10.1088/0004-637X/722/2/1030}, \href
  {http://adsabs.harvard.edu/abs/2010ApJ...722.1030W} {722, 1030}

\bibitem[\protect\citeauthoryear{{White} \& {Rees}}{{White} \&
  {Rees}}{1978}]{wr78}
{White} S.~D.~M.,  {Rees} M.~J.,  1978, \mndoi [\mnras]
  {10.1093/mnras/183.3.341}, \href
  {http://adsabs.harvard.edu/abs/1978MNRAS.183..341W} {183, 341}

\bibitem[\protect\citeauthoryear{{Willke} et~al.,}{{Willke}
  et~al.}{2002}]{waa02}
{Willke} B.,  et~al., 2002, \mndoi [Classical and Quantum Gravity]
  {10.1088/0264-9381/19/7/321}, \href
  {http://adsabs.harvard.edu/abs/2002CQGra..19.1377W} {19, 1377}

\bibitem[\protect\citeauthoryear{{Wyithe} \& {Loeb}}{{Wyithe} \&
  {Loeb}}{2003}]{wl03a}
{Wyithe} J.~S.~B.,  {Loeb} A.,  2003, \mndoi [ApJ] {10.1086/375187}, \href
  {http://adsabs.harvard.edu/abs/2003ApJ...590..691W} {590, 691}

\bibitem[\protect\citeauthoryear{{You} et~al.,}{{You} et~al.}{2007}]{yhc+07}
{You} X.~P.,  et~al., 2007, MNRAS, 378, 493

\bibitem[\protect\citeauthoryear{{Zhu} et~al.,}{{Zhu} et~al.}{2014}]{zhw14}
{Zhu} X.-J.,  et~al., 2014, \mnras, 444, 3709

\bibitem[\protect\citeauthoryear{{Zhu}, {Wen}, {Xiong}, {Xu}, {Wang},
  {Mohanty}, {Hobbs}  \& {Manchester}}{{Zhu} et~al.}{2016}]{zwx16}
{Zhu} X.-J.,  {Wen} L.,  {Xiong} J.,  {Xu} Y.,  {Wang} Y.,  {Mohanty} S.~D.,
  {Hobbs} G.,   {Manchester} R.~N.,  2016, \mndoi [\mnras]
  {10.1093/mnras/stw1446}, \href
  {http://adsabs.harvard.edu/abs/2016MNRAS.461.1317Z} {461, 1317}

\bibitem[\protect\citeauthoryear{{eLISA Consortium} et~al.,}{{eLISA Consortium}
  et~al.}{2013}]{eli13}
{eLISA Consortium} et~al., 2013, preprint (\mn@eprint {arXiv} {1305.5720})

\makeatother
\end{thebibliography}

\end{document}